\documentclass[12pt,a4paper]{article}
\usepackage{jheppub}
\usepackage{amssymb}
\usepackage{amsmath}
\usepackage{amsfonts}
\usepackage{mathrsfs}
\usepackage{mathtools}
\usepackage{graphicx}
\usepackage{enumitem}
\usepackage{xcolor}
\usepackage{dsfont}
\usepackage[utf8]{inputenc}
\usepackage{slashed}
\usepackage{amsmath}
\usepackage{bm}
\usepackage{bbm}
\usepackage{caption}	
\usepackage{enumerate}
\usepackage{braket}
\usepackage{comment}
\usepackage{ytableau}

\usepackage{chngcntr}
\usepackage{textcomp}
\usepackage{graphicx,longtable,afterpage}

\usepackage{floatrow}
\newfloatcommand{capbtabbox}{table}[][\FBwidth]

\def\one{{\,\hbox{1\kern-.8mm l}}}

\def\be{\begin{equation}}
\def\ee{\end{equation}}

\allowdisplaybreaks

\def\makeatletter{\catcode`\@=11}
\makeatletter
\def\mathbox#1{\hbox{$\m@th#1$}}%
\def\math@ccstyles#1#2#3#4#5#6#7{{\leavevmode
      \setbox0\mathbox{#6#7}%
      \setbox2\mathbox{#4#5}%
      \dimen@ #3%
      \baselineskip\z@\lineskiplimit#1\lineskip\z@
      \vbox{\ialign{##\crcr
             \hfil \kern #2\box2 \hfil\crcr
             \noalign{\kern\dimen@}%
             \hfil\box0\hfil\crcr}}}}
\def\mathaccstyles{\math@ccstyles\maxdimen}
\def\maththroughstyles{\math@ccstyles{-\maxdimen}}
\def\unity%
 {\maththroughstyles{.45\ht0}\z@\displaystyle {\mathchar"006C}\displaystyle 1}

\def\s{{\sigma}}

\def\beq{\begin{equation}}
\def\eeq{\end{equation}}
\newcommand{\bea}{\begin{eqnarray}}
\newcommand{\eea}{\end{eqnarray}}
\def\bal{\begin{align}}
\def\eal{\end{align}}

\newmuskip\pFqmuskip

\newcommand*\pFq[6][8]{%
  \begingroup 
  \pFqmuskip=#1mu\relax
  \mathchardef\normalcomma=\mathcode`,
  \mathcode`\,=\string"8000
  \begingroup\lccode`\~=`\,
  \lowercase{\endgroup\let~}\pFqcomma
  {}_{#2}F_{#3}{\left(\genfrac..{0pt}{}{#4}{#5}\Big | #6\right)}%
  \endgroup
}
\newcommand{\pFqcomma}{{\normalcomma}\mskip\pFqmuskip}

\def\be{\begin{equation}}
\def\ee{\end{equation}}
\def\bea{\begin{eqnarray}}
\def\eea{\end{eqnarray}}
\def\beal{\begin{equation}\begin{aligned}}
\def\eeal{\end{aligned}\end{equation}}

\def\eqn#1{Eq.~\eqref{#1}}
\def\cL{\mathcal{L}}

\def\eps{\epsilon}

\def\eps{\epsilon}

\def\be{\begin{equation}}
\def\ee{\end{equation}}
\def\bea{\begin{eqnarray}}
\def\eea{\end{eqnarray}}

\def\ep{\varepsilon}

\preprint{
\hfill CCTP-2025-05 \\ 
\mbox{}\hfill ITCP-IPP-2025/05 \\ 
\mbox{}\hfill QMUL-PH-25-12}

\title{
Searching for Kerr in string amplitudes}

\author{
F.~Alessio$^{\natural}$\;, P.~Di~Vecchia$^{\dagger,\nabla}$\;, M.~Firrotta$^{\flat}$\;, P.~Pichini\;$^{\ddagger}$}

\affiliation{
$^\natural$  \href{https://w3.lnf.infn.it}{INFN, Laboratori Nazionali di Frascati}, 00044 Frascati (RM), Italy  \\
~\\
$^\dagger$  \href{https://nordita.org}{NORDITA}, KTH Royal Institute of Technology and Stockholm University, \\ Hannes Alfv{\'{e}}ns v{\"{a}}g 12, SE-11419 Stockholm, Sweden  \\ 
~\\
$^\nabla$   \href{https://nbi.ku.dk/english/}{The Niels Bohr Institute}, Blegdamsvej 17, DK-2100, Copenhagen, Denmark\\
~\\
$^\flat$ \href{http://hep.physics.uoc.gr}{Crete Center for Theoretical Physics}, Institute for Theoretical and Computational Physics,
Department of Physics, Voutes University Campus,
GR-70013, Heraklion, Greece $ $ \\
~\\
$^\ddagger$  \href{https://www.qmul.ac.uk/spcs/ctp/}{Centre for Theoretical Physics}, Department of Physics and Astronomy,
Queen Mary University of London, Mile End Road, London E1 4NS, UK
}
\emailAdd{francesco.alessio@lnf.infn.it}\emailAdd{divecchi@nbi.dk}\emailAdd{mfirrotta@physics.uoc.gr}\emailAdd{p.pichini@qmul.ac.uk}
\abstract{We continue the approach of~\cite{Cangemi:2022abk}
 to attempt to reproduce the classical electromagnetic current of the $\sqrt{\mathrm{Kerr}}$ solution from the infinite-spin limit of a three-point amplitude with two higher-spin string states and a massless vector. We review the infinite-spin limit of three-point amplitudes along the leading Regge trajectory, which differ from $\sqrt{\mathrm{Kerr}}$, and we provide evidence that any sub-leading trajectory parallel to the leading one give the same result.
We extend the above investigation to the simplest sub-leading trajectory not parallel to the leading one, containing physical states with the first two harmonics. We compute all three-point amplitudes involving two such states and a massless vector, using both covariant and DDF formalism, and we extract the associated classical infinite-spin limit.
Inspired by the leading Regge case, where this limit reproduces the classical electromagnetic current sourced by a rigid rotating string, we derive new classical string solutions containing the first two harmonics and compare the currents they source to the infinite-spin amplitude of the two-harmonic state. 
We do not see a matching between the two approaches, and we find that the classical behaviour is instead reproduced by the amplitude involving coherent string states.
We conclude this paper by discussing common patterns between the different amplitudes we considered, and we conjecture how they can be generalised to reproduce $\sqrt{\mathrm{Kerr}}$.{\color{red}}
}

\date{\today}

\begin{document}

\maketitle

\hypersetup{pageanchor=true}

\setcounter{tocdepth}{2}


\section{Introduction}
\label{sect1}

The black holes that we observe at LIGO/VIRGO/KAGRA/GEO600 have a non-vanishing spin and are described in General Relativity (GR) by the Kerr black hole solution. In recent years, it was shown that Kerr observables can be obtained directly from the infinite-spin limit of scattering amplitudes with massive higher-spin states, and that the standard perturbative expansion for these amplitudes agrees with the post-Minkowskian expansion in GR.

The starting point of this construction is 
the linearised energy-momentum tensor of a Kerr black hole~\cite{Vines:2017hyw}. The energy-momentum tensor of a massive field coupled to gravity is given by the three-point amplitude of two massive particles and a graviton. Therefore, to reproduce Kerr from amplitudes, one has to first find a class of three-point amplitudes with two spin-$s$ particles and a graviton, for any value of $s$, and then show that it recovers the Kerr energy-momentum tensor for $s \to \infty$.

It turned out that the correct amplitudes are exactly the ones studied 
in Ref.~\cite{Arkani-Hamed:2017jhn} due to having the tamest possible high-energy behaviour and henceforth referred to as ``minimally-coupled''. Shortly after their first appearance, it was shown that they could be rewritten in terms of expectation values of the angular momentum operator instead of the standard polarisation vectors or spinors. In this new basis, these amplitudes are given by the exponential of the subleading soft operator~\cite{Guevara:2017csg,Guevara:2018wpp} and, in the infinite-spin limit, they recover the energy-momentum tensor of Kerr in terms of the covariant classical spin vector~\cite{Vaidya:2014kza,Arkani-Hamed:2017jhn,Cachazo:2017jef,Guevara:2017csg,Guevara:2018wpp,Chung:2018kqs,Johansson:2019dnu,Maybee:2019jus,Arkani-Hamed:2019ymq,Damgaard:2019lfh,Aoude:2020onz,Guevara:2020xjx,Chiodaroli:2021eug,Cangemi:2022bew,Cangemi:2023ysz,Cangemi:2023bpe,Gambino:2024uge,Bjerrum-Bohr:2025lpw}. Similarly, an analogous class of amplitudes with two equal spin-$s$ particles and a photon reproduces the electromagnetic current of the gauge-theory counterpart of Kerr black holes, known as the $\sqrt{\mathrm{Kerr}}$ solution~\cite{Arkani-Hamed:2019ymq}.

However, accessing higher-order observables to all orders in the black hole's spin requires a full understanding of the higher-point amplitudes for higher-spin particles that describe Kerr, starting with the four-point tree-level Compton amplitude with two spin-$s$ particles and two gravitons.

The first attempt at computing the correct Compton amplitude used factorisation constraints and recursion relations to try to derive it on-shell from the known three-point amplitudes~\cite{Arkani-Hamed:2017jhn,Johansson:2019dnu,Guevara:2018wpp,Chung:2018kqs,Aoude:2020onz,Falkowski:2020aso}. However, this approach lead to contact-term ambiguities.
To resolve this issue, a number of different avenues were explored. Some groups constructed generic effective field theories, or on-shell scattering amplitudes, directly in terms of classical spin vectors and tried to find sensible constraints to fix the free coefficients of the classical Compton amplitude~\cite{Guevara:2020xjx,Bern:2022kto,Aoude:2022trd,Aoude:2023vdk,Haddad:2023ylx,Bjerrum-Bohr:2023jau,Bjerrum-Bohr:2023iey,Vazquez-Holm:2025ztz}.
Others studied quantum field theory in the worldline formalism and found relations between black-hole amplitudes and symmetries of the worldline action~\cite{Jakobsen:2021zvh,Comberiati:2022cpm,Ben-Shahar:2023djm,Haddad:2024ebn}.
Another approach studied solutions to the Teukolsky equation, describing the propagation of gravitational waves on Kerr spacetime, and found a way to extract the Compton amplitude from it~\cite{Bautista:2021wfy,Bautista:2022wjf,Bautista:2023sdf,Bautista:2024agp}. 
Alternatively, other groups focussed on studying the massive higher-spin theories underlying the three-point amplitudes connected to Kerr~\cite{Aoude:2020mlg,Chiodaroli:2021eug,Cangemi:2022bew,Cangemi:2023bpe}. In particular, one of the authors of this work found a set of physical constraints which uniquely determines the correct three-point amplitudes, and used it to derive the Compton amplitude to all spin orders. 

All the approaches mentioned above have been able to obtain state-of-the-art classical observables for Kerr black holes and binary systems~\cite{Guevara:2019fsj,Chung:2019duq,Aoude:2020mlg,Chung:2020rrz,Damgaard:2022jem,Kosmopoulos:2021zoq,Chen:2021kxt,Jakobsen:2022fcj,Alessio:2022kwv,Aoude:2022thd,Menezes:2022tcs,Jakobsen:2022zsx,Chen:2022clh,Saketh:2022wap,Alessio:2023kgf,Bautista:2023szu,Jakobsen:2023ndj,Scheopner:2023rzp,Buonanno:2024vkx,Chen:2024mmm,Chen:2024bpf,Akpinar:2024meg,Bautista:2024agp,Bohnenblust:2024hkw,Gatica:2024mur,Akpinar:2025bkt,Jakobsen:2021lvp,Bautista:2021inx,Riva:2022fru,Heissenberg:2023uvo,DeAngelis:2023lvf,Brandhuber:2023hhl,Aoude:2023dui,Bohnenblust:2023qmy,Alessio:2025flu,Bohnenblust:2025gir,Akpinar:2025huz}. However, all the answers provided are either truncated to a finite spin order or still present some ambiguities, and the complete tree-level Compton amplitude for Kerr is still missing. Moreover, higher-order black-hole observables require knowledge of amplitudes at five points and above, which are still an open problem.

Studying higher-spin amplitudes in string theory provides a new perspective on this problem. Indeed, since string theory is a consistent quantum theory containing infinitely-many higher-spin states, it is natural to see if one can find the amplitudes describing Kerr within it~\cite{Cangemi:2022abk,Azevedo:2024rrf}. One advantage of this is that, assuming there is a string state that reproduces known Kerr amplitudes, all higher-point amplitudes will be uniquely determined by the state's vertex operator, thus providing a novel realisation of (long-distance) black-hole physics from strings. As a first step, one can consider the infinite-spin limit of all sorts of three-point tree-level string amplitudes to see what classical objects emerge from it. We follow the approach in Ref.~\cite{Cangemi:2022abk}, where it was shown that three-point amplitudes with two spin-$s$ leading Regge states and one massless boson, in the open and closed superstring, reproduce the electromagnetic current and energy momentum tensor sourced by known classical string solutions representing a rigid rotating string \cite{Ademollo:1974te}, in the $s \to \infty$ limit. However, leading-Regge amplitudes give a different result from Kerr, so to make further progress one needs to investigate subleading string  states.

It is easy to find the physical states at the few lowest levels, but at higher levels it becomes more subtle due to the complexity of solving physical-state constraints, and a general solution at an arbitrary level is not known.
An analysis of the structure of the physical states up to level ten is presented in Ref.~\cite{Pesando:2024lqa}. Moreover, recently a new framework has been constructed~\cite{Markou:2023ffh,Basile:2024uxn} which organises the physical states in trajectories,
starting not from the tachyon as in the case of the leading trajectory, but from some other physical state. One can, for example, start from the antisymmetric two-form state at the level $n=3$, described by a Young diagram with one column and two rows, and consider all physical states described by Young diagrams where the first row becomes longer and longer. In this way one gets all physical states lying on a trajectory parallel to the leading one. In this paper we will study this trajectory in more detail, and we will also discuss trajectories of states that are not parallel to the leading Regge one.

The physical states can also be described by the so-called DDF states~\cite{DelGiudice:1971yjh}. These form a complete set of states if $D=26$, and they have been successfully employed in many different contexts~\cite{Capolongo:2024gkg,Biswas:2024unn,Firrotta:2022cku,Biswas:2024mdu,Firrotta:2023wem,Bianchi:2023uby,Firrotta:2024fvi,Aldi:2019osr}.
The disadvantage of these states is that they provide scattering amplitudes with irreducible representations of $SO(D-2)$. It turns out, however, that if we restrict ourselves to the \textit{zero-depth} states defined in Ref.~\cite{Markou:2023ffh,Basile:2024uxn}, their three-point amplitude with a massless vector reproduces exactly the amplitude that one gets from the covariant formalism. This is one of the results of this paper.

In this work we begin the investigation of such subleading states in the open bosonic string. We start by reviewing the leading-Regge case and generalise the classical-limit technology developed in the $D=4$ case to arbitrary dimensions. We define a change of variables in the amplitude which replaces all massive polarisation tensors $\ep_i^a$ in favour of spin tensors $S^{ab}$ (as opposed to the spin vectors $S^a$ appearing in the four-dimensional case). Then, we take the infinite-spin limit to extract the classical result, which is expressed in terms of two modified Bessel functions of the first kind, $I_0$ and $I_1$, and for $D = 4$ it agrees with the result in Ref.~\cite{Cangemi:2022abk}. 

\begin{figure}[h!]
\centering
\includegraphics[scale=0.7]{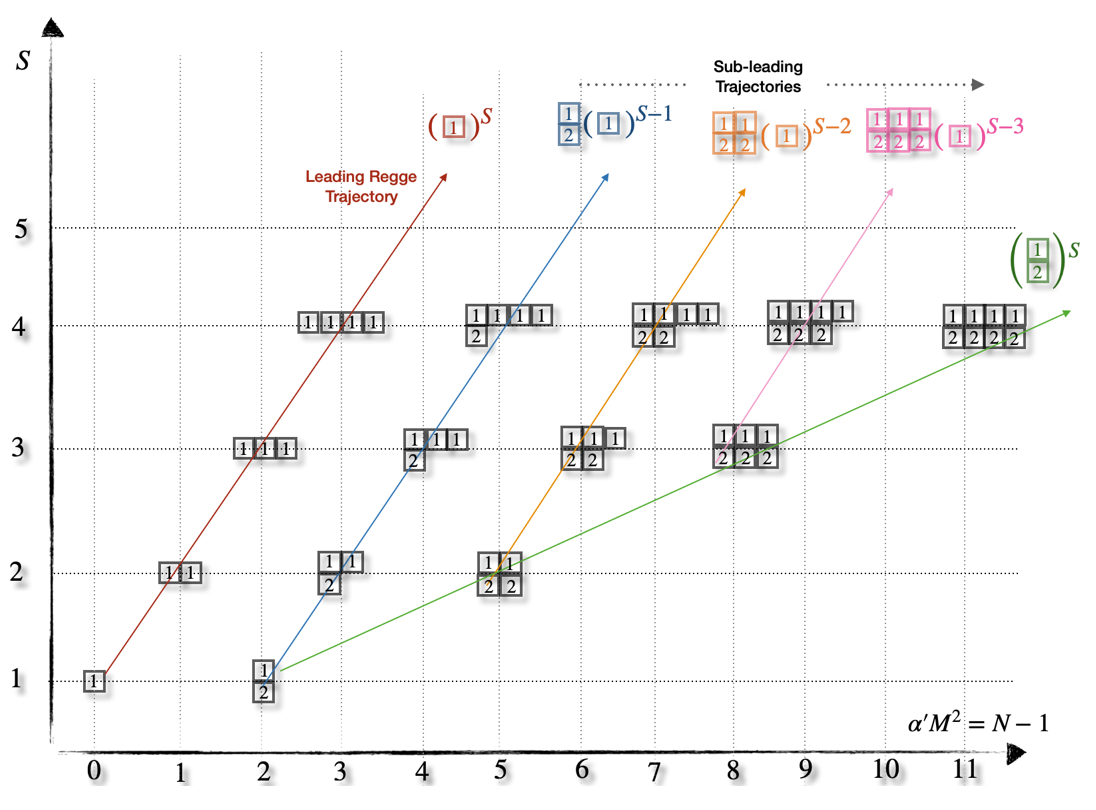}
\caption{The low-energy spectrum of the open bosonic string, organised in physical trajectories. In this work we study the leading Regge trajectory (red), the next-to-leading trajectory (blue) and the ``tilted trajectory'' (green).}
\label{fig:trajectories}
\end{figure}

Then we move on to consider subleading states. In this work, we restrict our attention to zero-depth states, defined as the lightest states for a given Young diagram in the string spectrum; for example, leading Regge states are zero-depth states for the single-row Young diagrams, see figure (\ref{fig:trajectories}). They have the important property of giving the same three-point amplitude by using both the DDF and the covariant formalism,

The first non-leading Regge zero-depth states we consider belong to the ``next-to-leading'' Regge trajectory, $\alpha' m^2=s+1$,  namely those states containing one copy of the string mode $a_{2}^{\dagger }$ and $s$ copies of the fundamental mode $a_{1}^{\dagger}$. A more precise definition is given in \eqn{stateNLR}. As shown in Fig.~\ref{fig:trajectories}, this trajectory is parallel to the leading-Regge one. We compute novel three-point amplitudes between two equal states from this trajectory and a massless vector. Then we compute its classical infinite-spin limit and we find that it agrees with the leading Regge result. This is not surprising and it can be understood as two parallel lines converging at infinity. From this calculation we believe that any trajectory with the same slope as the leading Regge trajectory will give the same infinite-spin result and therefore, if we hope to reproduce the $\sqrt{\mathrm{Kerr}}$ result, we need to study trajectories with a different slope.

The simplest zero-depth trajectory~\footnote{Here and in the following we use the word `` trajectory" also to mean a straight line containing physical states that is not necessarily parallel to what is normally called leading Regge trajectory and its daughters.} with a slope that is different from leading Regge is the one where each state contains $s$ copies of both the oscillators $a_{1}^\dagger$ and $a_{2}^\dagger$. Those states are discussed in more detail in \eqn{eq:gentwomodestate}, and they lie on  a  trajectory, $\alpha' m^2=3s-1$, displayed in Fig.~\ref{fig:trajectories} that we refer to as \textit{tilted trajectory}. We proceed to computing the three-point amplitudes between two such states and one massless vector, in arbitrary dimensions, making use of both covariant and DDF techniques \cite{Ademollo:1974kz, Bianchi:2019ywd, Firrotta:2024qel} and finding agreement between the two approaches. The resulting amplitudes, expressed nicely in terms of simple generalised Laguerre polynomials in \eqn{3pt2HaZetfac}, are one of the main results of this work. Interestingly, the amplitude for leading Regge states can also be recast in a similar form in terms of generalised Laguerre polynomials, as we show in \eqn{amplitudeLRs}. This provides a common structure for string amplitudes involving seemingly very different states, and therefore we believe it can be generalised to a broader class of states.

Then, we study the infinite-spin limit of these amplitudes.  
We will make use of the arbitrary-dimension classical limit since, as we will see, the states only exist in $D>4$ dimensions. Moreover, the classical limit was previously considered only for definite-spin states, i.e. representations with a single-row Young diagram. To repeat the analysis for this new trajectory, we generalised the classical-limit technology to the case of Young diagrams with more than one row. The result is given in \eqn{eq:infAtworow} in terms of the first four modified Bessel functions of the first kind and therefore, it is different  from the leading Regge one~\eqref{eq:infspinA3LR}.

Since in the leading Regge case, the infinite-spin amplitude  corresponds to the electromagnetic current of a classical solution of a rigid rotating string, one could wonder which is the classical string solution that reproduces the infinite-spin result for the tilted trajectory. This motivated us to study also classical string solutions with two oscillator modes $a_1$ and $a_2$. However, we find that the electromagnetic current obtained from such solutions is very different from the infinite-spin limit of the amplitude for the tilted trajectory states.

In order to understand this discrepancy, we turn our attention to string coherent states, which are expected to reproduce classical observables and have been studied in the literature as an alternative approach to the infinite-spin limit~\cite{Aoude:2021oqj,Cangemi:2023ysz,Cangemi:2023bpe}. We construct a coherent state $|\psi\rangle$ written  in terms of the string harmonic oscillators that satisfies the following condition,
\begin{equation}
\label{stringcoherent}
    \langle \psi|L_n |\psi\rangle=0,
\end{equation}
for any integer $n$. The expectation value $\bra{\psi} X^\mu \ket{\psi}$ automatically satisfies the string equations of motion and constraints, showing that there is a one-to-one correspondence between coherent states and classical string solutions. We show that the amplitudes involving states in \eqref{stringcoherent} always reproduce the electromagnetic current of the associated solution. We conclude that coherent states are the correct states to consider when trying to reproduce classical observables from amplitudes, and even though in the case of leading Regge they agree with the infinite-spin limit along a string trajectory, this is not true in general, and the tilted trajectory provides one example of this.

The paper is organised as follows. In Sect.~\ref{sect2} we review the calculation of the three-point amplitude involving two equal states of the leading Regge trajectory and a photon, and we compute the classical limit in arbitrary dimensions. In Sect.~\ref{sect:hook} we compute the three-point amplitude involving two states on the next-to-leading Regge trajectory and a photon and we argue that in the $s\rightarrow\infty$ limit it yields the same result of the leading Regge one. In Sect.~\ref{sect4.1} we define the zero-depth states with a Young diagram given by two rows of the same length. We compute the three-point amplitude with two such states and a photon and compute its classical limit. In Sect.~\ref{sect5} we construct classical string solutions containing one and two harmonics. In the one-harmonic case, we reproduce the known matching to the infinite-spin limit of the leading Regge amplitudes. In the two-harmonics case, we show that there is no four-dimensional solution and we construct a new explicit solution in $D>4$ dimensions. We also compute its electromagnetic current and show that it does not agree with the classical limit of the quantum amplitude discussed in \ref{sect4.2}. In Sect.~\ref{sect6} we study the coherent state formalism and show that it is in one-to-one correspondence with classical solutions. We compute three-point amplitudes for coherent states with one and two harmonics and compare them to the infinite-spin limit of quantum amplitudes.

\section{Massless state coupled to leading Regge states}
\label{sect2}

In the present Section we review the computation of the 3-point amplitude involving two spin-$s$ leading Regge states and one massless state in open bosonic string theory.\footnote{ We consider only one of the two orderings and we omit to write the corresponding Chan-Paton factor.} We also give a novel representation of such amplitude, obtained from an appropriately-defined generating function, which is suitable for computing the $s\to\infty$ limit and thereby extracting the classical multipole expansion. 

\subsection{Three-point amplitude of the leading Regge trajectory}
\label{sect2.1}
We begin by considering a spin-$s$ state from the leading Regge trajectory of the open bosonic string,
\begin{equation}
\label{nonfactLR}
   |\psi_{\,\mathrm{LR}}^{(s)}(p)\rangle:= {1\over \sqrt{s!}} \ep_{b_1 \dots b_s}a_1^{\dagger b_1} \dots a_1^{\dagger b_s} \ket{0,p} ,
\end{equation}
with $b_{k}=0\dots D-1$ and mass shell condition $\alpha'm^{2}=s-1$. The oscillators $a_n^{\dagger }$, with $n\in(0,\infty)$, are creation operators for each string mode.\footnote{Here the oscillator variables are normalised such that $[a_n^{b},a_m^{\dagger\,c}]=\delta_{nm}\eta^{ac}$.} The polarization tensor $\ep_{a_1 \dots a_s}$ is a fully-symmetric rank-$s$ tensor which satisfies transversality and tracelessness, 
\be\label{PhyCondLR}
p^{a_1} \ep_{(a_1 \dots a_s)}=0\,,\qquad \eta^{a_1 a_2} \ep_{(a_1 \dots a_s)}=0\,.
\ee
The Young diagram associated to states on the leading-Regge trajectory is the following:

\begin{equation}
\label{eq:youngLR}
\ytableausetup
 {mathmode, boxframe=normal, boxsize=2em}
{\tiny{\begin{ytableau}
 {} & & & \none[\dots] & {} \\
\end{ytableau}}}
\quad .
\end{equation}
The degrees of freedom of the polarization tensor~\eqref{PhyCondLR} are
\be
\begin{pmatrix} s+D-2 \\ s\end{pmatrix}-\begin{pmatrix}s+D-4\\s-2 \end{pmatrix}\,.
\ee
In $D=4$ the tensor has $2s+1$ free components, reproducing the expected counting for integer spin-$s$ representations of the $SO(3)$ massive little group.
In order for the state in \eqref{nonfactLR} to have unit norm, we impose 
$  \overline{\varepsilon}_{a_{1}\dots a_{s}}\varepsilon^{a_{1}\dots a_{s}}=1$.\\
Instead of computing the amplitude involving the state \eqref{nonfactLR}, one can introduce a factorised version\footnote{Strictly speaking, the state $\ket{\Psi_{\,\mathrm{LR}}^{(s)}}$ has $D-1$ degrees of freedom, but it has the same symmetries of \eqref{nonfactLR} and satisfies the same constraints.} of such state
\be
\label{factLR}
   \ket{\Psi_{\,\mathrm{LR}}^{(s)}(p)} := {1\over \sqrt{s!}}(\ep\cdot a^{\dagger }_1)^s \ket{0,p} ,
\ee
where $\ep$ satisfies $p{\cdot}\ep= 0$ and $\ep^2 = 0$ according to \eqref{PhyCondLR} and we further normalise it as $\ep\cdot\bar\ep=1.$ Any observable for the state $\ket{\Psi_{\mathrm{LR}}^{(s)}}$ can be converted into the corresponding one for  $\ket{\psi_{\,\mathrm{LR}}^{(s)}}$ via the relation\begin{equation}
\label{replrule}
    \ket{\psi_{\,\mathrm{LR}}^{(s)}(p)} ={1\over s!}\ep_{a_1 \dots a_s} \frac{\partial}{\partial \ep_{a_1}} \dots \frac{\partial}{\partial \ep_{a_s}} \ket{\Psi_{\,\mathrm{LR}}^{(s)}(p)} .
\end{equation}
The three-point amplitude for two leading Regge states with spin $s$ and one massless vector state is given by 
\be\label{3ptLR}
{\cal A}_{\,\mathrm{LR}}^{(s)}:=\bra{\Psi_{\,\mathrm{LR}}^{(s)}(p_{1})}\oint {dz\over 4\alpha'\pi i\,z}\, {\epsilon}{\cdot}i\partial \hat{X}\,e^{ik{\cdot}\hat{X}}(z)\ket{\Psi_{\,\mathrm{LR}}^{(s)}(p_{2})},
\ee
where $\epsilon$ is the polarization vector of the massless state transverse to its light-like momentum $k$, $\epsilon{\cdot}k=0$ and $\epsilon^2=k^{2}=0$. For the string field $\hat{X}$, we take the mode expansion
\begin{equation}
\label{eq:modexp}
    \hat{X}^a = 2\alpha'\hat{p}^a \tau + i\sqrt{2\alpha'} \sum_{n=1}^\infty \frac{1}{\sqrt{n}} \cos{n\sigma} (a_n^a e^{-i n\tau}-a_n^{\dagger a} e^{+i n\tau}),
\end{equation}
where $\hat{p}$ is the momentum operator, and $a_n^{\dagger a}$ and $a_n^a$ are the creation and annihilation operators for the $n$-th string mode, satisfying $[a_n^a,a_n^{\dagger b}] = \eta^{a b}$. Using the coherent-state method described in appendix \ref{GenffLR}, we can derive the generating function of the amplitudes \eqref{3ptLR}, defined as
\be\label{LRgenFF}
\mathcal{A}^{(\mathrm{gen})}_{\,\mathrm{LR}}:=\langle p_{1}| e^{c_{1}\varepsilon_{1}{\cdot}a_{1}} \,\oint{dz\over 4\alpha'\pi i\,z}\, {\epsilon}{\cdot}i\partial \hat{X}\,e^{ik{\cdot}\hat{X}}(z)\, e^{c_{2}\varepsilon_{2}{\cdot}a^{\dagger}_{1}}|p_{2}\rangle\,,
\ee
in terms of which the spin-$s$ amplitude is obtained as
\be
{\cal A}_{\,\mathrm{LR}}^{(s)}={1\over \sqrt{s!}}\left({\partial\over \partial c_{1}}\right)^{s}{1\over \sqrt{s!}}\left({\partial\over \partial c_{2}}\right)^{s}{\cal A}_{\mathrm{LR}}^{(\mathrm{gen})}\big|_{c_1\mapsto0\atop c_2\mapsto 0}.
\ee
Introducing the variables 
\be
\label{eq:xyLR}
y:=\varepsilon_1\cdot\varepsilon_2,\qquad x:=\frac{2\alpha '\varepsilon_1\cdot k\varepsilon_2\cdot k}{y},
\ee
we find 
\begin{align}
\label{eq:GfinalLR}
\mathcal{A}^{(\mathrm{gen})}_{\,\mathrm{LR}}=\bigg(-\epsilon\cdot p_1
\,J_{0}(2\sqrt{c_{1}c_{2}xy})+c_{1}c_{2}\,\varepsilon_1{\cdot} f{\cdot}\varepsilon_2\frac{J_{1}(2\sqrt{c_{1}c_{2}xy})}{\sqrt{c_{1}c_{2}xy}}\bigg)e^{c_{1}c_{2}y},
\end{align}
where $f^{ab}=k^{[a}\eps^{b]}$ is the field strength of the massless state and $J_n(x)$ are Bessel functions of the first kind, given by
\begin{align}
J_n(x):=\sum_{r=0}^{\infty}\frac{(-)^r}{r!(r+n)!}\left(\frac{x}{2}\right)^{2r+n}.
\end{align}
It is worth noticing that the previous equation can be rewritten in terms of generalised Laguerre polynomials $L_n^{(\alpha)}(x)$ as 
\begin{align}\label{amplitudeS}
\mathcal{A}^{(\mathrm{gen})}_{\,\mathrm{LR}}=\sum_{k=0}^{\infty}y^k\bigg(-\frac{(c_{1}c_{2})^k}{k!}\epsilon{\cdot}p_1\,L_{k}^{(0)}(x)+\frac{(c_{1}c_{2})^{k+1}}{(k+1)!}\varepsilon_1{\cdot}f{\cdot}\varepsilon_2L_{k§}^{(1)}(x)\bigg),
\end{align}
where $L_n^{(\alpha)}(x)$ are defined as
\begin{align}
L_n^{(\alpha)}(x):=\sum_{k=0}^n(-1)^k\binom{n+\alpha}{n-k}\frac{x^k}{k!}\,.
\end{align}
The amplitude \eqref{3ptLR} can be immediately extracted from \eqref{amplitudeS}
and it simply reads
\begin{align}
\label{amplitudeLRs}
\mathcal{A}_{\,\mathrm{LR}}^{(s)}=-\epsilon{\cdot}p_1 \,y^s L_s^{(0)}(x)+\varepsilon_1{\cdot}f{\cdot}\varepsilon_2\,y^{s-1}L_{s-1}^{(1)}(x).
\end{align}

\subsection{Spin multipole expansion and classical limit}
\label{sect2.2}
Having computed the quantum amplitude, we can proceed to rewrite it into expectation values of the spin operator $\hat{S}^{ab}$, to then take the infinite-spin limit and extract the classical spin-multipole expansion. The exact procedure to achieve this is described in detail in Ref.~\cite{Cangemi:2022abk}. Here we will extend it to $D$ generic dimensions.

Given the string spin operator $\hat{J}^{ab}$ \footnote{We define symmetrization and antisymmetrization as $A^{(a}B^{b)}=A^{a}B^{b}+A^{b}B^{a}$ and $A^{[a}B^{b]}=A^{a}B^{b}-A^{b}B^{a}$.}
\begin{align}
\label{spinoperator}
\hat{J}^{ab}:=-i\sum_{n=1}^{\infty}a^{\dagger[a}_na_n^{b]},
\end{align}
the spin operator $\hat{S}^{ab}$ is defined as the projection of $\hat{J}^{ab}$ perpendicular to the reference momentum $p^{b}$,
\begin{align}
  \hat{S}^{ab}:=P^a{}_c \, P^b{}_d\hat{J}^{cd},\qquad P^a{}_b:=\delta^a_b+\frac{p^ap_b}{m^2}.
\end{align}
 In particular, following appendix \ref{LRexVals}, we compute the expectation values\footnote{We denote $\langle \Psi_{\,\mathrm{LR}}^{(s)}|\hat{\mathcal{O}}|\Psi_{\,\mathrm{LR}}^{(s)}\rangle:=\langle \hat{\mathcal{O}}\rangle_{\,\mathrm{LR}}^{(s)}$ for any observable $\hat{\cal O}$.},
\begin{subequations}
\begin{align}\label{SLR}
&\langle \hat{S}^{ab}\rangle_{\,\mathrm{LR}}^{(s)}= i s\,\varepsilon^{[a}\bar{\varepsilon}^{b]},\\
&\label{SLR1}\langle \hat{S}^{ac}\hat{S}_c{}^b\rangle_{\,\mathrm{LR}}^{(s)}=-s(s{-}1)\varepsilon^{(a}\bar{\varepsilon}^{b)}{-}s\big[P^{a b}{+}(D{-}3)\bar{\varepsilon}^{a}\varepsilon^b\big],\\& \label{SLR2}\langle \hat{S}^{ab}\hat{S}_{b a}\rangle_{\,\mathrm{LR}}^{(s)}=-2s(s+D-3).
\end{align}
\end{subequations}

Notice that in $D=4$ we recover the familiar result $\langle \hat{S}^{2}\rangle_{\mathrm{LR}}^{(s)}=s(s+1)$, where $\hat{S}^{a}=-\frac{1}{2m}\epsilon^{abcd}p_{b}\hat{S}_{cd}$ is the spin vector. In the classical, large-$s$ limit we correctly get the factorisation
\begin{align}
    \langle \hat{S}^{ac}\hat{S}_{c}{}^{b}\rangle_{\mathrm{LR}}^{(s)}\stackrel{s\to\infty}{\sim}\langle \hat{S}^{ac}\rangle_{\mathrm{LR}}^{(s)}\langle \hat{S}_{c}{}^{b}\rangle_{\mathrm{LR}}^{(s)},
\end{align}
as we should. Now we can focus of the amplitude \eqref{amplitudeLRs}. The first step is to rewrite the three-point amplitude in terms of a single massive momentum $p_1$ and a single massive polarisation vector $\ep_1$. This is because the amplitude describes a single classical massive object, but it depends on two massive external legs with momenta $p_1$ and $p_2$ and polarisation vectors $\ep_1$ and $\ep_2$. This is achieved by expressing $p_2$ and $\varepsilon_2$ via a Lorentz boost acting on $p_1$ and $\varepsilon_1$,
\begin{align}
\label{boost}
\varepsilon_2^{a}=\bar{\varepsilon}_1^{a}+\frac{k\cdot\bar{\varepsilon}_1}{m^2}\Big(p_1^{a}+\frac{k^{a}}{2}\Big).
\end{align}
Notice that, on the three-point kinematics $p_1+p_2+k=0$, $p_i^2 = -m^2$ and $k^2 = p_1\cdot k = 0$, the above formula preserves the transversality $\varepsilon_2\cdot p_2=0$. Using \eqref{boost} we get
\begin{align}
y=1+\frac{|\varepsilon_1\cdot k|^2}{2m^2},\qquad x=2\alpha'\frac{|\varepsilon_1\cdot k|^2}{y},\\ \varepsilon_1\cdot f\cdot \varepsilon_2=\varepsilon_1\cdot f\cdot \bar{\varepsilon}_1-\epsilon\cdot p_1\frac{|\varepsilon_1\cdot k|^2}{m^2},
\end{align}
where we used $\varepsilon_2\cdot\bar{\varepsilon}_2=\varepsilon_1\cdot\bar\varepsilon_1=1$. Having done this boost, the amplitude now depends on $s$ powers of the polarisation vector $\ep_1$ and $s$ powers of its conjugate $\bar{\ep}_1$.

The next step is to change variables from the massive polarisation vectors to expectation values of the spin operator $\hat{S}^{ab}$. In principle, for a spin-$s$ state, the amplitude will be a function of $\langle \hat{S}^{a_1b_1}\dots \hat{S}^{a_nb_n}\rangle_{\mathrm{LR}}^{(s)}$ with $0\leq n\leq s$. However, working with the factorised state in \eqn{factLR}, it is enough to consider $\langle\hat{S}^{ab}\rangle_{\mathrm{LR}}^{(s)}$ and $\langle\hat{S}^{ac}\hat{S}_{c}{}^{b}\rangle_{\mathrm{LR}}^{(s)}$, and arbitrary powers of them. This is because, as can be seen in \eqn{SLR} and \eqn{SLR1}, each pair $\ep^a \bar\ep^b$ can be rewritten in terms of the projector $P^{ab}$ and those two structures. Note that this alternative approach does not affect the classical $s \to \infty$ limit, due to the property 
\begin{align}
\langle \hat{S}^{a_1 b_1}\dots \hat{S}^{a_n b_n}\rangle_{\mathrm{LR}}{}^{(s)}\stackrel{s\to\infty}{\sim}\langle \hat{S}^{a_1 b_1}\rangle_{\mathrm{LR}}^{(s)}\dots\langle\hat{S}^{a_n b_n}\rangle_{\mathrm{LR}}^{(s)}.
\end{align}
Explicitly, we can use \eqref{SLR} and \eqref{SLR1} and obtain~\footnote{We adopt the notation $u_a\langle\hat{S}^{ab}\rangle^{(s)}v_b\equiv u\cdot\langle \hat{S}\rangle^{(s)}\cdot v$.} 
\begin{align}
|\varepsilon_1\cdot k|^2=-\frac{k\cdot\langle \hat{S}\cdot \hat{S}\rangle_{\mathrm{LR}}^{(s)}\cdot k}{s(2s+D-5)},\qquad \varepsilon_1\cdot f\cdot\bar{\varepsilon}_1=-\frac{i}{2s}\langle \hat{S}^{ab}\rangle_{\mathrm{LR}}^{(s)} f_{ab},
\end{align}
and in the large $s$ limit
the amplitude in \eqref{amplitudeLRs} becomes
\begin{multline}
\mathcal{A}^{(s)}_{\mathrm{LR}}\stackrel{s\to\infty}{=}-\epsilon\cdot p_1\,L_s^{(0)}\bigg(-\frac{k\cdot\langle \hat{S}\cdot \hat{S}\rangle_{\mathrm{LR}}^{(s)}\cdot k}{s\, m^2}\bigg)\\-\frac{i}{2s}f_{ab}\langle \hat{S}^{ab}\rangle_{\mathrm{LR}}^{(s)} L_{s-1}^{(1)}\bigg(-\frac{k\cdot\langle \hat{S}\cdot \hat{S}\rangle_{\mathrm{LR}}^{(s)}\cdot k}{s\, m^2}\bigg),
\end{multline}
where we used the leading Regge trajectory relation $\alpha' = s/m^2$ for large $s$. We can now apply the asymptotic formula,
\begin{align}
\label{asymptL}
L_s^{(\alpha)}(x/s)\stackrel{s\to\infty}{\sim}J_{\alpha}(2\sqrt{x})(s/\sqrt{x})^{\alpha},
\end{align}
and 
therefore we get
\begin{align}
\label{eq:infspinA3LRbesselJ}
\mathcal{A}^{(\infty)}_{\mathrm{LR}}= -\epsilon\cdot p_1\,J_0(2\sqrt{-k\cdot a\cdot a\cdot k})-i\,m\frac{f^{ab} a_{ab}}{2}\frac{J_1(2\sqrt{-k\cdot a\cdot a\cdot k})}{\sqrt{-k\cdot a\cdot a\cdot k}},
\end{align}
where we denoted
\begin{align}
\label{classicalJ}
\lim_{s\to\infty}\langle \hat{S}^{ab}\rangle_{\mathrm{LR}}^{(s)}=S^{ab}=:m \,a^{ab},\qquad \lim_{s\to\infty}\frac{k\cdot\langle \hat{S}\cdot \hat{S}\rangle_{\mathrm{LR}}^{(s)}\cdot k}{m^2}=k\cdot a\cdot a\cdot k,
\end{align}
where we introduced the classical tensor  $a^{ab}$ which we refer to as the \textit{ring radius tensor}, in analogy to Ref.~\cite{Cangemi:2022abk}. Notice that, although \eqref{eq:infspinA3LRbesselJ} is the $s\rightarrow\infty$ limit of an amplitude, its structure is similar to the one appearing in the generating function in \eqref{eq:GfinalLR}. As we discuss later in Sect.~\ref{sect6}, this is not an accident and  it is related to the fact that \eqref{eq:GfinalLR} can actually be interpreted as an amplitude for coherent states. Using $J_n(ix)=i^n I_n(x)$, the result can be expressed in terms of modified Bessel functions of the first kind as 
\begin{align}
\label{eq:infspinA3LR}
\mathcal{A}^{(\infty)}_{\mathrm{LR}}=-\epsilon\cdot p_1 I_0(2\sqrt{k\cdot a\cdot a\cdot k})-i m\frac{f^{ab}a_{ab}}{2\sqrt{k\cdot a\cdot a\cdot k}}I_1(2\sqrt{k\cdot a\cdot a\cdot k}).\end{align}

In four dimensions we can introduce the ring radius vector $a$ as 
\begin{align}
a^{ab}=\frac{1}{m}\epsilon^{abcd}p_{1\,c}a_{d}.
\end{align}
In this way we get
\begin{align}
k\cdot a\cdot a\cdot k=-\frac{1}{m^2}\epsilon^{a}(k,p_1,a)\epsilon_a(k,p_1,a)=(k\cdot a)^2,
\end{align}
where we have taken into account of the fact that $p_1\cdot a=p_1\cdot k=0$. Using also $f^{ab}a_{ab}=\frac{2}{m}\epsilon(k,\epsilon,p_1,a)$ 
 and $\epsilon(k,\epsilon,p_1,a)\epsilon(k,\epsilon,p_1,a)=
-(a\cdot k )^2(p_{1}\cdot\epsilon)^2$ 
we can recover the result of \cite{Cangemi:2022abk},\footnote{In order to obtain \eqref{eq:classALR} we assumed that $\epsilon$ has positive helicity. In the negative helicity case the odd-in-spin multipoles change sign.}
\begin{align}
\label{eq:classALR}
\mathcal{A}_{\mathrm{LR}}^{(\infty)}=-\epsilon\cdot p_1\Big(I_0(2a\cdot k)+I_1(2a\cdot k)\Big).
\end{align}
It is instructive to compare the above result to the classical three-point amplitude for $\sqrt{\mathrm{Kerr}}$, that we write as
\begin{align}
\label{eq:Kerr}
\mathcal{A}^{(\infty)}_{\sqrt{\mathrm{Kerr}}}=-\epsilon\cdot p_1\,e^{a\cdot k}=-\epsilon\cdot p_1\Big(I_0(a\cdot k)+2\sum_{n=1}^{\infty}I_n(a\cdot k)\Big).
\end{align}
We notice that the two amplitudes above agree only up to the dipole, the latter being universal in the string, and they start to be different from the quadrupole. In particular, \eqn{eq:Kerr}, contains all the Bessel functions with non-negative integer index, whereas \eqref{eq:classALR} only the first two. In other words, following e.g. Ref.~\cite{Alaverdian:2025jtw}, one can write a general three-point amplitude in terms of  Wilson coefficients $C_n$, as
\begin{align}
\mathcal{A}^{(\infty)}=-\epsilon\cdot p_1\Big(\sum_{n=0}^{\infty}\frac{C_{2n}}{(2n)!}(k\cdot a)^{2n}+\sum_{n=0}^{\infty}\frac{C_{2n+1}}{(2n+1)!}(k\cdot a)^{2n+1}\Big),
\end{align}
and we get for $\sqrt{\mathrm{Kerr}}$ and leading-Regge
\begin{align}
C^{\sqrt{\mathrm{Kerr}}}_{2n}=1,\qquad C^{\sqrt{\mathrm{Kerr}}}_{2n+1}=1,\qquad C^{\mathrm{LR}}_{2n}=\binom{2n}{n},\qquad C^{\mathrm{LR}}_{2n+1}=\binom{2n+1}{n},
\end{align}
respectively.

It is also worth noticing that Ref.~\cite{Bianchi:2024shc} showed that a similar structure involving Bessel functions appears when computing the classical stress-energy tensor of Myers-Perry black holes.

In the next sections we show that more Bessel functions can be generated by considering states involving also the second string harmonic. 

\section{Massless state coupled to next-to-leading Regge states}
\label{sect:hook}
In this Section we consider states lying on the ``next-to-leading" Regge trajectory and their coupling with a massless state. The Young tableaux associated to these states is
\begin{equation}
\label{eq:younghook}
{\tiny{\ytableausetup
 {mathmode, boxframe=normal, boxsize=2em}
\begin{ytableau}
 {} & & & \none[\dots] & {} \\ 
 {}     {} \\
\end{ytableau}}}
\quad,
\end{equation}
and they can be taken to be, for $s\geq 1$
\begin{align}
\label{stateNLR}
|\Psi^{(s)}_{\mathrm{NLR}}(p)\rangle=N(\varepsilon,\tilde{\varepsilon},s) (\varepsilon\cdot a_1^{\dagger
})^{s-1}(\zeta_{b_1b_2}a_1^{b_1\dagger}a_2^{b_2\dagger})|0,p\rangle,\quad\zeta_{b_1 b_2}:=\frac{1}{2}\varepsilon_{[b_1}\tilde{\varepsilon}_{b_2]}.
\end{align}
It is easy to show that the state \eqref{stateNLR} satisfies the physical-state conditions, given by
\begin{equation}
\label{eq:qmphyscond}
    (L_{n}-\delta_{n,0})|\Psi^{(s)}_{\mathrm{NLR}}(p)\rangle = 0 ,
\end{equation}
for $n\geq 0$, provided that the vectors $\varepsilon$ and $\tilde{\varepsilon}$ satisfy
\begin{align}
\label{phystateNLR}
p\cdot \varepsilon=p\cdot\tilde{\varepsilon}=0,\qquad \varepsilon^2=\tilde{\varepsilon}^2=\varepsilon\cdot\tilde{\varepsilon}=0.
\end{align}
Note that $L_n$ are the generators of the Virasoro algebra, related to the oscillators in \eqn{eq:modexp} by
\begin{equation}
    L_{n\neq0} = \frac{1}{2}\sum_{k\in\mathbb{Z}} \hat\alpha_k\cdot\hat\alpha_{n-k} \quad , \qquad L_0 = \frac{1}{2} \hat\alpha_0^2 + \sum_{k=1}^\infty \hat\alpha_{-k}\cdot\hat\alpha_{k} ,
\end{equation}
where $\hat\alpha_0 = \sqrt{2\alpha'} \, \hat{p}$, and $\hat\alpha_{n} = \sqrt{n}a_n$ and $\hat\alpha_{-n} = a_{-n}^{\dagger }$ for $n > 0$. 

Notice that the first state of the trajectory in \eqref{stateNLR}, for $s=1$, is a massive two-form, which in $D=4$ can be dualised to a massive vector having spin one. The mass-shell condition for these states implies $\alpha'm^2=s+1$, where $s$ in $D=4$ can indeed be identified as the spin quantum number. Such trajectory of states, as depicted in Fig.~\ref{fig:trajectories}, is parallel to the leading-Regge one. 

We are interested in computing the three-point amplitude involving two states  \eqref{stateNLR} and a massless vector,

\begin{align}
\mathcal{A}_{\mathrm{NLR}}^{(s)}:=\langle\Psi_{\mathrm{NLR}}^{(s)}(p_1)|\oint\frac{dz}{4\alpha' \pi i z}\epsilon\cdot i\partial \hat{X}\, e^{ik\cdot \hat{X}}(z)|\Psi_{\mathrm{NLR}}^{(s)}(p_2)
\rangle
,\end{align}
that can be obtained from the generating function 
\begin{align}
\label{defgenNLR}
\mathcal{A}^{(\mathrm{gen})}_{\mathrm{NLR}}:=\zeta_{ab}^{1}\zeta^2_{cd}\langle p_1|e^{c_1\varepsilon_1\cdot a_1}a_1^aa_2^b\oint\frac{dz}{4\alpha'\pi i z}\eps\cdot i\partial \hat{X}\,e^{ik\cdot \hat{X}}(z)e^{c_2\varepsilon_2\cdot a^{\dagger}_1}a^{\dagger c}_1a^{\dagger d}_2|p_2\rangle,
\end{align}
as
\begin{align}
\label{eq:amp}
\mathcal{A}^{(s)}_{\mathrm{NLR}}:=N(\varepsilon_1,\tilde{\varepsilon}_1,s)N(\varepsilon_2,\tilde{\varepsilon}_2,s)\bigg(\frac{\partial}{\partial c_1}\bigg)^{s-1}\bigg(\frac{\partial}{\partial c_2}\bigg)^{s-1}\mathcal{A}^{(\mathrm{gen})}_{\mathrm{NLR}}\big|_{c_1\mapsto0\atop c_2\mapsto 0}.
\end{align}
As described in appendix \ref{appNLR}, the generating function can be computed straightforwardly and it yields, similarly to the leading Regge case, an expression involving Bessel functions of the first kind. However, the presence of the oscillator $a_2^{\dagger}$ generates the additional Bessel function $J_2(x)$ which was absent in \eqref{eq:GfinalLR}. The explicit result for the generating function is 
\begin{align}
\label{eq:genfNLR}
\mathcal{A}^{(\mathrm{gen})}_{\mathrm{NLR}}=\bigg(-\eps\cdot p_1\sum_{l=0}^1 h_l\frac{J_l(2\sqrt{c_1c_2xy})}{(c_1c_2xy)^{\frac{l}{2}}}+f_{ab}\sum_{l=0}^2d_l^{ab}\frac{J_l(2\sqrt{c_1c_2xy})}{(c_1c_2xy)^{\frac{l}{2}}}\bigg)e^{c_1c_2y},
\end{align}
with coefficients
\be
h_l=\sum_{k=0}^{1}h_l^{(k)}(c_1 c_2)^k\,,\qquad d_l^{ab}=\sum_{k=0}^2 d_l^{ab\,(k)}(c_1c_2)^k,
\ee
where
\be
h_0^{(0)}=-\mathrm{Tr}(\zeta_{1}\cdot \zeta_2) + \alpha'3 k\cdot \zeta_1\cdot \zeta_2\cdot k \,, \quad h_0^{(1)}=-\varepsilon_2\cdot \zeta_1\cdot\zeta_2\cdot\varepsilon_1-\varepsilon_2\cdot \zeta_1\cdot k\varepsilon_1\cdot \zeta_2\cdot k,
\ee
\be
h_1^{(0)}=0\,,\quad h_1^{(1)}=2\alpha' (\varepsilon_1\cdot k\varepsilon_2\cdot \zeta_1\cdot \zeta_2\cdot k+\varepsilon_2\cdot k k\cdot \zeta_1\cdot \zeta_2\cdot\varepsilon_1),
\ee
and
\be
d_0^{ab\,(0)}=-2 (\zeta_1\cdot \zeta_2)^{ab}\,,\quad d_0^{ab\,(1)}=-(\varepsilon_2\cdot \zeta_1)^b(\varepsilon_1\cdot \zeta_2)^a\,,\quad d_0^{ab\,(2)}=0,
\ee
\be
d_1^{ab\,(0)}=0\,,\quad d_1^{ab\,(2)}=\varepsilon_2\cdot \zeta_1\cdot \zeta_2\cdot\varepsilon_1\varepsilon_{2}^{a}\varepsilon_{1}^{b} + \alpha'\varepsilon_2\cdot \zeta_1\cdot k\varepsilon_1\cdot \zeta_2\cdot k\varepsilon_2^{a}\varepsilon_1^b,
\ee
\be
d_1^{ab\,(1)}=2\,\mathrm{Tr}(\zeta_1\cdot \zeta_2)\varepsilon_{2}^{a}\varepsilon_{1}^{b}-\frac{\alpha'}{2}(6k\cdot \zeta_1\cdot \zeta_2\cdot k\varepsilon_{2}^{a}\varepsilon_{1}^{b}-k\cdot\varepsilon_2 \zeta_1^{ab}\varepsilon_1\cdot \zeta_2\cdot k+k\cdot\varepsilon_1 \zeta_2^{ab}\varepsilon_2\cdot \zeta_1\cdot k),
\ee
\be
d_2^{ab\,(0)}=d_2^{ab\,(1)}=0\,,\quad d_2^{ab\,(2)}=-2\alpha'\varepsilon_1\cdot k\varepsilon_2\cdot k\mathrm{Tr}(\zeta_1\cdot \zeta_2)\varepsilon_2^{a}\varepsilon_1^b,
\ee
where, as for leading Regge,
\begin{align}
y:=\varepsilon_1\cdot\varepsilon_2,\qquad x:=\frac{2\alpha'\varepsilon_1\cdot k\varepsilon_2\cdot k}{y}.
\end{align}
The generating function can be rewritten as
\begin{multline}
\label{eq:genfunctNLR2}
\mathcal{A}^{(\mathrm{gen)}}_{\mathrm{NLR}}=\bigg[-\eps\cdot p_1\sum_{l,k=0}^1 h_l^{(k)}\sum_{n\geq 0}\frac{(c_1c_2)^{n+k}y^n}{(n+l)!}L_n^{(l)}(x)\\+f_{ab}\sum_{l,k=0}^2d_l^{ab\,(k)}\sum_{n\geq 0}\frac{(c_1c_2 )^{n+k}y^n}{(n+l)!}L_n^{(l)}(x)\bigg],
\end{multline}
where we used the formula
\begin{align}
e^{y}\frac{J_{l}(2\sqrt{xy})}{(xy)^{\frac{l}{2}}}=\sum_{n\geq 0}\frac{y^n}{(n+l)!}L_n^{(l)}(x).
\end{align}
The amplitude can now easily be computed and it is given by 
\begin{multline}
\label{eq:amplNLR}
\mathcal{A}_{\mathrm{NLR}}^{(s)}=p(s)y^{s-1}\bigg[-\eps\cdot p_1\sum_{l=0}^1\sum_{k=0}^{\mathrm{min}(1,s-1)}\frac{h_l^{(k)}y^{-k}}{(s-1+l-k)!}L_{s-1-k}^{(l)}(x)\\+f_{ab}\sum_{l=0}^2\sum_{k=0}^{\mathrm{min}(2,s-1)}\frac{d_l^{ab\,(k)}y^{-k}}{(s-1+l-k)!}L_{s-1-k}^{(l)}(x)\bigg],
\end{multline}
where $p(s)$ is an overall factor depending on the normalisations that we do not need to specify here. It can be shown that, going through the same procedure discussed in Sect.~\ref{sect2.2}, in the infinite-$s$ limit, only the coefficients $c_0^{(0)}$ and $d_1^{ab\,(1)}$ survive and, once all the quantities are expressed in terms of the classical spin structures, one finds again that, in four-dimensions
\begin{align}
\label{eq:classamplNLR}
\mathcal{A}_{\mathrm{NLR}}^{(\infty)}=-\epsilon\cdot p_1\Big(I_0(2a\cdot k)+I_1(2a\cdot k)\Big),
\end{align}
exactly as for leading-Regge. This result is telling us that, even if for finite-$s$ the amplitudes for states lying on the leading and next-to-leading Regge trajectories are different, in the infinite-spin limit they give exactly the same result. This should not be entirely surprising, because the contribution of the oscillator $a^{\dagger}_2$ in the state \eqref{stateNLR}, which makes it different with respect to the one in \eqref{factLR}, is actually irrelevant as $s\rightarrow\infty$. From this simple argument it would follow that all the trajectories parallel to the leading-Regge one, obtained by increasing only the number of oscillators $a^{\dagger}_1$, will always yield the same result in \eqref{eq:classamplNLR} in the infinite-spin limit. In order to obtain something different it is therefore necessary to consider trajectories where not only the number of $a_1^{\dagger}$ oscillators increases as $s\to\infty$, but also that of other oscillators $a_{n>1}^{\dagger}$. The easiest of these states, proportional to $(a_1^{\dagger}a_2^{\dagger})^s$ will be considered in the next section.

\section{Massless state coupled to tilted trajectory states}
\label{Sect:2HS}
As anticipated, in this Section we will focus on the state
\begin{equation}
\label{eq:gentwomodestate}
\ket{\psi^{(s)}_{\mathrm{2h}}(p)} :={1\over s!} T_{a_1 \dots a_s ; b_1 \dots b_s}a_{1}^{\dagger a_1}\dots a_{1}^{\dagger a_s} a_{2}^{\dagger b_1}\dots a_{2}^{\dagger b_s} \ket{0,p},
\end{equation}
where we take $T_{a_1 \dots a_s ; b_1 \dots b_s}\bar{T}^{a_1 \dots a_s ; b_1 \dots b_s}=1$
so that $\langle{\psi^{(s)}_{\mathrm{2h}}}\ket{\psi^{(s)}_{\mathrm{2h}}}=1$. 
Without loss of generality we assume $T_{a_1 \dots a_s ; b_1 \dots b_s} = T_{(a_1 \dots a_s) ; b_1 \dots b_s}/s! = T_{a_1 \dots a_s ; (b_1 \dots b_s)}/s!$ since these are the symmetries imposed by the oscillators $a_n^{\dagger}$. 
\label{sect4}
\subsection{Physical conditions of the tilted trajectory}
\label{sect4.1}
The physical-state conditions on the state~\eqref{eq:gentwomodestate} are, as before in \eqn{eq:qmphyscond},
\begin{equation}
    \label{eq:virasoro} (L_{n} - \delta_{n,0}) \ket{\psi^{(s)}_{\mathrm{2h}}(p)} = 0 ,
\end{equation}
for $n \geq 0$. Due to the Virasoro algebra, it is enough to consider just the action of $L_1$ and $L_2$, that can be written as
\begin{align}
\begin{split}
    &L_1=\sqrt{2\alpha'} p\cdot a_1 +\sqrt{2} a_1^\dagger \cdot a_2+ \dots,\\ 
    &L_2= \sqrt{2\alpha'}\sqrt{2}\, p \cdot a_2 + \frac{1}{2} a_1 \cdot a_1 + \dots ,
\end{split}
\end{align}
where we have omitted terms that automatically annihilate the state in \eqn{eq:gentwomodestate}.
Imposing $L_1\ket{\psi_s} = L_2\ket{\psi_s} = 0$ we find the constraints
\begin{align}
\label{eq:twomodestatephyscond}
\begin{split}
    &p^{a_1} T_{a_1 \dots a_s ; b_1 \dots b_s} = p^{b_1} T_{a_1 \dots a_s ; b_1 \dots b_s} = 0 ,\\
    &T_{(a_1 \dots a_s ; b_1) \dots b_s} = 0 ,\\
    &\eta^{a_1 a_2} T_{a_1 \dots a_s ; b_1 \dots b_s} = 0 ,\\
    &s! \,T_{a_1 \dots a_s ; b_1 \dots b_s} = T_{(a_1 \dots a_s) ; b_1 \dots b_s} = T_{a_1 \dots a_s ; (b_1 \dots b_s)} ,
\end{split}
\end{align}
where we included the last relation, assumed previously, for convenience. 
The equation in the second line together with the first equation in the first line comes from the action of $L_1$, while the equation in the third line together with the second equation in the first line comes from the action of $L_2$.

The relations in \eqn{eq:twomodestatephyscond} define a mixed-symmetry irreducible representation of the little group $SO(D-1)$, namely the one described by the Young diagram
\begin{equation}
\label{eq:youngtworow}
\ytableausetup
 {mathmode, boxframe=normal, boxsize=2em}
{\tiny{\begin{ytableau}
 {} & & & \none[\dots] & {} \\ 
 {} & & & \none[\dots] & {} \\
\end{ytableau}}}
\quad,
\end{equation}
where each row has $s$ boxes. Note that \eqn{eq:twomodestatephyscond} implies additional relations, such as
\begin{align}
\label{eq:twomodestateextracond}
\begin{split}
    &T_{a_1 \dots (a_s ; b_1 \dots b_s)} = 0 ,\\
    &\eta^{a_1 b_1} T_{a_1 \dots a_s ; b_1 \dots b_s} = \eta^{b_1 b_2} T_{a_1 \dots a_s ; b_1 \dots b_s} = 0 .
\end{split}
\end{align}
It is possible to count the number of degrees of freedom in the tensor $T_{a_1 \dots a_s ; b_1 \dots b_s}$ based on the constraints listed above. Following the discussion in Ref.~\cite{663b51c5-158f-325a-916b-03575665ab5f} about representations of the orthogonal group, the general result in $D$ dimensions is\footnote{Actually, the results in \cite{663b51c5-158f-325a-916b-03575665ab5f} lead to an additional factor of $(2s+D-4)$ with respect to \eqn{eq:twomodestateextracond}. We believe this is a typo in \cite{663b51c5-158f-325a-916b-03575665ab5f}.}
\begin{equation}
\label{eq:twomodeDOF}
\frac{(s+D-5)!(s+D-6)!(2s+D-3)(2s+D-4)(2s+D-5)}{(D-3)!(D-5)!s!(s+1)!} ,
\end{equation}
for the representation in \eqn{eq:youngtworow}.

Similarly to the leading-Regge case, we also consider the factorised version of \eqref{eq:gentwomodestate}, which can be written as
\begin{equation}
\label{eq:tworowfactorepsilon}
\ket{\Psi_{\mathrm{2h}}^{(s)}(p)} := N(\zeta,s)(\zeta_{a b} a^{\dagger a}_1 a_2^{\dagger b})^s \ket{0,p} ,\quad \zeta_{a b}:={1\over 2}\ep_{[a}\tilde{\ep}_{b]}.
\end{equation}

Note that the physical-state conditions now imply
$p^a\zeta_{ab} = 0$ and $(\zeta\cdot\zeta)_{a b} = 0$ and they translate again into \eqref{phystateNLR}, $p\cdot\ep = p\cdot\tilde{\ep} = 0$ and $\ep^2 = \tilde{\ep}^2 = \ep\cdot\tilde{\ep} = 0$. This implies that the normalisation is
\begin{align}
N(\zeta,s)=\frac{1}{\sqrt{s!(s+1)!(n_\zeta)^s}},\qquad  n_\zeta:=-\frac{\mathrm{Tr}(\zeta\cdot\bar{\zeta})}{2},
\end{align}
with
\begin{align}
    \mathrm{Tr\,(\zeta\cdot\bar{\zeta})}=\frac{1}{2}(\varepsilon\cdot\bar{\tilde{\varepsilon}}\tilde{\varepsilon}\cdot\bar{\varepsilon}-|\varepsilon|^2|\tilde{\varepsilon}|^2).
\end{align}
The above normalisation ensures that $\langle{\Psi_{\mathrm{2h}}^{(s)}}\ket{\Psi_{\mathrm{2h}}^{(s)}}=1$. Moreover, for simplicity we choose to normalise $\ep$ and $\tilde\ep$ such that $n_\zeta = 1$.

In the rest of this work, we will use the state $\ket{\Psi_{\mathrm{2h}}^{(s)}}$ to compute amplitudes and extract their classical limit. The corresponding results in terms of the state $\ket{\psi_{\mathrm{2h}}^{(s)}}$ can always be recovered using the relation
\begin{equation}
\label{eq:zetafromepsilon}
    \ket{\psi_{\mathrm{2h}}^{(s)}} = \frac{2^s}{s!(s+1)!\sqrt{s+1}} T_{c_1\dots c_s d_1\dots d_s} \frac{\partial^2}{\partial \ep_{[c_1} \partial \tilde{\ep}_{d_1]}} \dots \frac{\partial^2}{\partial \ep_{[c_s} \partial \tilde{\ep}_{d_s]}} \ket{\Psi_{\mathrm{2h}}^{(s)}},
\end{equation}
 and acting on the amplitude with the above differential operator for each particle belonging to the tilted trajectory.

\subsection{Three-point amplitude of the tilted trajectory}
\label{sect4.2}
In this Section we present the computation of the three-point amplitude involving one massless state and two states on the tilted trajectory. We give explicit expressions of amplitudes with excited states in terms of both $\zeta_i$ and $(\varepsilon_i,\tilde{\varepsilon}_i)$. Similarly to the leading Regge case, we first compute the generating function
(see appendix \ref{GenffTwH} for a detailed derivation) and then extract the amplitudes.

The amplitude we are interested in can be obtained by computing the following correlation function  
\be\label{3pt2HaZet}
{\cal A}_{\mathrm{2h}}^{(s)}:=\bra{\Psi_{\mathrm{2h}}^{(s)}(p_{1})}\oint {dz\over 4\alpha'\pi i\,z}\, {\epsilon}{\cdot}i\partial \hat{X}\,e^{ik{\cdot}\hat{X}}(z)\ket{\Psi_{\mathrm{2h}}^{(s)}(p_{2})}\,,
\ee
that can be extracted from its generating function, defined as 
\be\label{3pt2HaZetgen}
{\cal A}_{\mathrm{2h}}^{(\mathrm{gen})}:=\bra{p_{1}}e^{c_1\zeta^1_{b_1 b_2} a^{b_1}_1 a_2^{b_2}}\oint{dz\over 4\alpha'\pi i\,z}\, {\epsilon}{\cdot}i\partial \hat{X}\,e^{ik{\cdot}\hat{X}}(z)\,e^{c_2\zeta^2_{b_1 b_2} a^{\dagger b_1}_1 a_2^{\dagger b_2}}\ket{p_2}\,,
\ee
using
\be\label{amplProj2h}
{\cal A}_{\mathrm{2h}}^{(s)}=N(\zeta_1,s)N(\zeta_2,s)\left({\partial\over \partial c_1} \right)^s \left({\partial\over \partial c_2} \right)^s {\cal A}_{\mathrm{2h}}^{(\mathrm{gen})}\big|_{c_1\mapsto0\atop c_2\mapsto 0}.
\ee
Following appendix \ref{GenffTwH}, we find the following simple expression:
\be\label{genf2h}
{\cal A}_{\mathrm{2h}}^{(\mathrm{gen})}=-\bigg(\epsilon{\cdot} p_1-2 f_{ab}\left({c_1 c_2 M\over \mathbbm{1}+ c_1 c_2 M}\right)^{ab}\bigg)e^{-[\log(\mathbbm{1}+c_1 c_2 M)]^{a}{}_a + 3\alpha' k{\cdot}\left( {c_1 c_2 M\over \mathbbm{1}+c_1 c_2 M} \right){\cdot}k}\,,
\ee
where we used
\be
[\log(\mathbbm{1}+c_1 c_2 M)]^{a}{}_a= - \sum_{n=1}^\infty {(-)^n\over n} (c_1 c_2)^n  (M^n)^{a}{}_a,
\ee
and
\be
\left({c_1 c_2 M\over \mathbbm{1}+ c_1c_2 M}\right)^{ab}= - \sum_{n=1}^\infty (-)^n (c_1 c_2)^n (M^n)^{ab}\,,
\ee
with the matrix $M^n$ defined as
\begin{align}
\label{generators}
(M^n)^{ab}:=\underbrace{(\zeta^2\cdot\zeta^1\dots\zeta^2\cdot\zeta^1}_{n})^{ab},\qquad(M^0)^{ab}:=\eta^{ab}.
\end{align}
Further introducing
\begin{align}
C_\ell := \frac{1}{\ell} (M^\ell)^{a}{}_a - 3\alpha' k\cdot (M^\ell) \cdot k\,,\qquad B_{2m}:= \prod_{i=1}^m\sum_{n_i=0}^\infty (-1)^{in_i}\frac{C_i^{n_i}}{n_i!} \delta_{\sum_{i=1}^{} i n_i,m} \, ,
\label{O3P9}  
\end{align}
one can extract the amplitude using \eqref{amplProj2h}, obtaining 
\begin{align}
\label{eq:amplitudeB}
\mathcal{A}^{(s)}_{\mathrm{2h}}=-(s!)^2N(s,\zeta_1)N(s,\zeta_2)\big(\epsilon\cdot p_1 B_{2s}+2f_{ab}\sum_{n=0}^s(-1)^n(M^n)^{ab}B_{2(s-n)}\big).
\end{align}

We quote below the explicit result of the amplitudes for $s=1$ (massive two-form) and $s=2$:
\begin{align}
{\cal A}_{\mathrm{2h}}^{(1)}=&-\epsilon\cdot p_1(M^{a}{}_a - 3\alpha'k{\cdot}M{\cdot}k) - 2 f_{ab}M^{ab}\\
\nonumber{\cal A}_{\mathrm{2h}}^{(2)}=&-\epsilon{\cdot}p_1  \Bigg[\frac{( M^a{}_a)^2}{2} + \frac{(M^2)^a{}_a}{2}
 - 3\alpha' ( k{\cdot}M{\cdot}k M^a{}_a +
 k{\cdot}M^2{\cdot}k)+\frac{9}{2} (\alpha'k{\cdot}M{\cdot}k)^2\Bigg]\\
 &-2 f_{ab}[(M^2)^{ab}+M^{ab}M^{c}{}_c-3  M^{ab}\alpha'k{\cdot}M{\cdot}k] \,.
 \end{align}

Note that, so far, the generating function in \eqref{genf2h} and the amplitude in \eqref{eq:amplitudeB} are valid for any $\zeta_i$ satisfying the physical constraints. Choosing now $\zeta_i$ as in \eqref{eq:tworowfactorepsilon} they simplify and, defining 
\begin{align}
&(\zeta_2\cdot \zeta_1)^{ab}=\frac{1}{4}(\varepsilon_2^a\tilde{\varepsilon}_1^b\tilde{\varepsilon}_2\cdot\varepsilon_1-\varepsilon_2^a\varepsilon_1^b\tilde{\varepsilon}_2\cdot\tilde{\varepsilon}_1-\tilde{\varepsilon}_2^a\tilde{\varepsilon}_1^b\varepsilon_2\cdot\varepsilon_1+\tilde{\varepsilon}_2^a\varepsilon_1^b\varepsilon_2\cdot\tilde{\varepsilon}_1),\\&\mathrm{Tr\,(\zeta_2\cdot\zeta_1)}=\frac{1}{2}(\varepsilon_2\cdot\tilde{\varepsilon}_1\tilde{\varepsilon}_2\cdot\varepsilon_1-\varepsilon_2\cdot\varepsilon_1\tilde{\varepsilon}_2\cdot\tilde{\varepsilon}_1),
\end{align}
one has the following relation
\begin{align}
\label{eq:factpower}
(M^n)^{ab}=\frac{1}{2^{n-1}}\big(\mathrm{Tr}(\zeta_2\cdot\zeta_1)\big)^{n-1}(\zeta_2\cdot\zeta§_1)^{ab},\qquad n>0\,.
\end{align}
This allows to resum the generating function as
\begin{align}\label{genfunc2hfacComp}
{\cal A}_{\mathrm{2h}}^{(\mathrm{gen})}=-\bigg(\frac{\epsilon\cdot p_1}{(1-c_1 c_2 v)^2}-f_{ab}(\zeta_2\cdot\zeta_1)^{[ab]}\frac{c_1c_2}{(1-c_1 c_2 v)^3}\bigg)e^{-{c_1 c_2 v\,u \over 1-c_1 c_2 v}}\, ,
\end{align}
where we have introduced the quantities 
\begin{align}v:=-\frac{\mathrm{Tr}(\zeta_1\cdot\zeta_2)}{2},\qquad
u:=-3\alpha'\frac{k\cdot\zeta_1\cdot\zeta_2\cdot k}{v},\qquad 
\end{align}
which play the same role as $x$ and $y$ introduced in \eqref{eq:xyLR} for leading-Regge.
Interestingly, we notice that also here, using the generating function for the generalised Laguerre polynomials,
\begin{align}
    \sum_{k=0}^{\infty}L_k^{(\alpha)}(x)y^k=\frac{1}{(1-y)^{\alpha+1}}e^{-\frac{yx}{1-y}},
\end{align}
the expression \eqref{genfunc2hfacComp} can be can be represented as
\begin{align}
{\cal A}_{\mathrm{2h}}^{(\mathrm{gen})}=-
\sum_{k=0}^{\infty}v^k\bigg((c_1 c_2)^k\epsilon\cdot p_1L_{k}^{(1)}(u)-(c_1 c_2)^{k+1}f_{ab}(\zeta_2\cdot\zeta_1)^{[ab]}L_{k}^{(2)}(u)\bigg),\end{align}
and using \eqref{amplProj2h}, the result of the corresponding three-point amplitude is  
\begin{align}
\label{3pt2HaZetfac}
{\cal A}_{\mathrm{2h}}^{(s)}=-\frac{v^s}{s+1}\Big(\epsilon\cdot p_1 \, L_{s}^{(1)}(u)-f_{ab}(\zeta_2\cdot\zeta_1)^{[ab]}\,v^{-1}L_{s-1}^{(2)}(u)\Big).
\end{align}
It is worth remarking that the amplitude in \eqref{3pt2HaZetfac} has a similar structure to the one found for the leading-Regge case, given in \eqref{eq:classALR}. In fact, they both contain only two generalised Laguerre polynomials, but with different superscript indices. 
Therefore, a natural conjecture for the three-point amplitude of two physical states of the form $(\zeta_{b_1\dots b_n}a_1^{b_1\dagger} \dots a_n^{b_n\dagger})^s\ket{0}$ and a massless vector is 
\begin{align}
\mathcal{A}^{(s)}_n\sim-\epsilon\cdot p_1 \, L_{s}^{(n-1)}\left(X_n\right)+f_{ab}\langle S^{ab}\rangle^{(s)}_n\,L_{s-1}^{(n)}\left(X_n\right),
\end{align}
where $X_1 := x$ in \eqn{eq:classALR}, $X_2 := u$ and $X_{n>2}$ needs to be defined appropriately to generalise the $n = 1,2$ cases. Moreover, we expect a large class of three-point string amplitudes to be expressed in terms of generalised Laguerre polynomials, and we have already provided a nontrivial example in \eqn{eq:amplNLR}. Understanding the details of this structure in full generality is an interesting direction which we leave for future work.

\subsection{Spin multipole expansion and classical limit}
\label{sect4.3}
In analogy with the leading Regge case, we can represent the amplitude \eqref{3pt2HaZetfac} in terms of a spin-multipole expansion and study its classical limit. 
 Similarly, we present the analysis for the factorised state in \eqref{eq:tworowfactorepsilon}, but using the mapping in \eqref{eq:zetafromepsilon}, it can be extended to the most general physical polarization tensor.
We start by computing the spin and spin-squared expectation values. Following appendix \ref{TwHexVals}, we find 
\begin{subequations}
\begin{align}
\label{eq:S}
&\langle\hat{S}^{ab}\rangle_{\mathrm{2h}}^{(s)}=-is(\zeta\cdot\bar{\zeta})^{[ab]}.,\\&
\label{eq:S2}
\langle \hat{S}^{ac}\hat{S}_{c}{}^{b}\rangle_{\mathrm{2h}}^{(s)}=(\zeta\cdot\bar{\zeta})^{ba}(D+s-5)+(\zeta\cdot\bar{\zeta})^{ab}(s-2)-2P^{ab},
\\&\langle \hat{S}^{ab}\hat{S}_{b a} \rangle_{\mathrm{2h}}^{(s)}=P_{ab}\langle \hat{S}^{ac}\hat{S}_{c}{}^{b}\rangle_{\mathrm{2h}}^{(s)}=-4s(s+D-4).
\end{align}
\end{subequations}
Notice that in $D=4$ one would get $\langle\hat{S}^2\rangle_{\mathrm{2h}}^{(s)}=-2s^2$ in terms of the spin vector.  
 They correctly satisfy the factorisation property 
\begin{align}
\langle \hat{S}^{ac}\hat{S}_c{}^{b}\rangle_{\mathrm{2h}}^{(s)}\stackrel{s\to\infty}{\sim}\langle\hat{S}^{ac}\rangle_{\mathrm{2h}}^{(s)}\langle\hat{S}_{c}{}^{b}\rangle_{\mathrm{2h}}^{(s)}.
\end{align}
Now, in order to express the amplitude in terms of $\langle \hat{S}^{ab}\rangle_{\mathrm{2h}}^{(s)}$ and $\langle \hat{S}^{ac}\hat{S}_c{}^b\rangle_{\mathrm{2h}}^{(s)}$ we need to compute the action of the boost in \eqref{boost}  on the quantum amplitude in \eqref{eq:amplitudeB}. Using \eqref{eq:tworowfactorepsilon} we find that  $\zeta_2^{ab}$ transforms as follows
\begin{align}
\zeta_{2}^{ab}=\bar\zeta_1^{ab}-\frac{k_{c}\bar\zeta_1^{c[a}}{m_1^2}\bigg(p_1^{b]}+\frac{k^{b]}}{2}\bigg).
\end{align}
Using this equation together with three-point kinematics yields
\begin{align}
&f_{ab}(\zeta_2\cdot\zeta_1)^{ab}=f_{ab}(\bar{\zeta}_1\cdot \zeta_1)^{ab}+\frac{\epsilon\cdot p_1}{m^2}k\cdot\zeta_1\cdot\bar{\zeta}_1\cdot k,\\
&\mathrm{Tr}(\zeta_1\cdot\zeta_2)=\mathrm{Tr}(\zeta_1\cdot\bar\zeta_1)+\frac{1}{m^2}k\cdot\zeta_1\cdot\bar\zeta_1\cdot k,\\&
k\cdot\zeta_1\cdot\zeta_2\cdot k=k\cdot\zeta_1\cdot\bar\zeta_1\cdot k.
\end{align}
 Let us now turn to the classical limit. From now on we set $\zeta_1\equiv\zeta$ to simplify the notation.
Notice that from \eqref{eq:S} and \eqref{eq:S2} we have
\begin{align}
k\cdot\zeta\cdot\bar\zeta\cdot k=\frac{k\cdot\langle\hat{S}\cdot\hat{S}\rangle_{\mathrm{2h}}^{(s)}\cdot k}{s(D+2s-7)},\qquad
f_{ab}(\zeta\cdot\bar{\zeta})^{ab}=\frac{i}{s}f_{ab}\langle\hat{S}^{ab}\rangle_{\mathrm{2h}}^{(s)}.
\end{align}
As a consequence, for large values of $s$, we have $v\sim1$ and we get (notice that the dependence on $D$ drops out in the $s\rightarrow \infty$ limit)
\begin{multline}
\mathcal{A}^{(s)}_{\mathrm{2h}}\stackrel{s\to\infty}{=}-\frac{1}{s+1}\bigg[\epsilon\cdot p_1L_s^{(1)}\bigg(-\frac{9}{2}\frac{k\cdot\langle \hat{S}\cdot\hat{S}\rangle_{\mathrm{2h}}^{(s)}\cdot k}{s\,m^2}\bigg)\\+\frac{i}{s}f_{ab}\langle\hat{S}^{ab}\rangle_{\mathrm{2h}}^{(s)} L_{s-1}^{(2)}\bigg(-\frac{9}{2}\frac{k\cdot\langle \hat{S}\cdot\hat{S}\rangle_{\mathrm{2h}}^{(s)}\cdot k}{s\,m^2}\bigg)\bigg],
\end{multline}
where we used the specific trajectory $\alpha'm^2\sim 3s$. Using again the asymptotic formula \eqref{asymptL} gives 
\begin{align}
\mathcal{A}^{(\infty)}_{\mathrm{2h}}=-\epsilon\cdot p_1 \frac{I_1(3\sqrt{2}\sqrt{k\cdot a\cdot a\cdot k})}{\frac{3}{\sqrt{2}}\sqrt{k\cdot a\cdot a\cdot k}}-imf_{ab}a^{ab}\frac{I_2(3\sqrt{2}\sqrt{k\cdot a\cdot a\cdot k})}{\frac{9}{2}k\cdot a\cdot a\cdot k},
\end{align}
where we used the same definitions as in \eqref{classicalJ} for the classical angular momentum tensor. The previous equation
can equivalently be written as
\begin{align}
\label{eq:infAtworow}
\nonumber\mathcal{A}_{\mathrm{2h}}^{(\infty)}=&-\epsilon\cdot p_1\Big(I_0(3\sqrt{2}\sqrt{k\cdot a\cdot a\cdot k})-I_2(3\sqrt{2}\sqrt{k\cdot a\cdot a\cdot k})\Big)\\&-i\frac{f_{ab}a^{ab}}{3\sqrt{2}\sqrt{k\cdot a\cdot a\cdot k}}\Big(I_1(3\sqrt{2}\sqrt{k\cdot a\cdot a\cdot k})-I_3(3\sqrt{2}\sqrt{k\cdot a\cdot a\cdot k})\Big).
\end{align} 
We can also specialise the above formula to $D = 4$. Note that the state~\eqref{eq:gentwomodestate} for $s>1$ has zero degrees of freedom in four dimensions, as can be computed from \eqn{eq:twomodeDOF}. However, a sensible $D= 4$ amplitude can be obtained via Kaluza-Klein compactification of a higher-dimensional state. In practice, we realise the compactification by assuming that the vectors $\ep_i$ lie in the extended four dimensions, as well as all momenta, whereas the vectors $\tilde{\ep}_i$ lie in the compact dimensions. This introduces the additional relations
\begin{equation}
    \tilde{\ep}_i \cdot \ep_j = \tilde{\ep}_i \cdot p_j = \tilde{\ep}_i\cdot k = 0 ,
\end{equation}
and it leads to the following classical amplitude,
\begin{align}
\label{eq:classATR}
\mathcal{A}^{(\infty)}_{\mathrm{2h}}\big|_{\substack{comp.\\D=4}}=-\epsilon\cdot p_1\Big((I_0(3\sqrt{2}a{\cdot}k)-I_2(3\sqrt{2}a{\cdot}k))+\frac{\sqrt{2}}{3}(I_1(3\sqrt{2}a{\cdot}k)-I_3  (3\sqrt{2}a{\cdot}k))\Big).
\end{align}

\section{Classical string solutions}
\label{sect5}

In this Section we will compare the infinite-spin amplitudes discussed above and the electromagnetic currents of classical string solutions. If only the fundamental mode $\alpha_1$ is turned on, the classical current reproduces the amplitude~\eqref{eq:classALR} obtained from the leading Regge trajectory, which was first found in Ref.~\cite{Cangemi:2022abk} and we will briefly review it. In the case of two modes turned on, namely $\alpha_1$ and $\alpha_2$, we will see that the classical current does not reproduce the amplitude~\eqref{eq:classATR} for the two-mode state considered in this work. Instead, it agrees with the amplitude obtained from string coherent states, as we will discuss in detail.

\subsection{Generic classical amplitude}
\label{sect5.1}
All classical string solutions described below are derived from the conformal gauge Lagrangian,
\begin{equation}
\cL = \frac{1}{4\pi \alpha'}\Big( {\dot{X}}^2 - (X')^2 \Big) ,
    \label{BB1}
\end{equation}
where $X^a(\tau,\sigma)$ is the target-space position of the string for each choice of the worldsheet parameters $\tau$ and $\sigma$, and ${\dot{X}_a}= \frac{\partial X_a}{\partial \tau} $ and $ X'_a = \frac{\partial X_a}{\partial \sigma} $. Solutions also satisfy the Virasoro constraints,
\begin{equation}
\label{eq:virasoro1}
({\dot{X}}\pm X')^2=0 .
\end{equation}
An open string classical solution satisfies the following equation of motion and boundary conditions,
\begin{align}
&X''_a-\ddot{X}_a=0~~~;~~ X'_a =0\quad \text{for} \quad \sigma=0, \pi \quad ,
    \label{BB3}
\end{align}
together with the constraints in \eqref{eq:virasoro1}. It is convenient to work in the proper time gauge, where the time is given by
\begin{equation}
X^0 =2\alpha' m \tau ,
\label{BB4}
\end{equation}
and in the center-of-mass frame $p^a = (m,\vec{0})$. The most general solution has form
\begin{align}
\label{eq:genclassX}
X^a = 2\alpha'p^a \tau + i\sqrt{2\alpha'} \sum_{n\neq 0}\frac{\alpha_n^a}{n} e^{-i n\tau}\cos{n\sigma}, 
\end{align}
where the parameters $\alpha_n^a$ are constrained by \eqn{eq:virasoro1}, which we can rewrite as
\begin{equation}
\label{eq:virasoro2}
    L_n = \frac{1}{2} \sum_{k\in\mathbb{Z}} \alpha_k\cdot\alpha_{n-k} = 0 .
\end{equation}
This is the classical counterpart of the physical-state conditions~\eqref{eq:qmphyscond}.

Following Ref.~\cite{Cangemi:2022abk}, we can define a classical amplitude $\mathcal{A}^{\text{cl}}$  by placing a charge $Q$ at one endpoint of the string, $\sigma = 0$, and computing the interaction between the electromagnetic current in momentum space, $j^a$, and an on-shell photon $\epsilon^a$,
\begin{equation}
\label{eq:classA3}
\mathcal{A}^{\text{cl}}
:= \frac{Q}{4\pi \alpha'} \int_0^{2\pi}\,d\tau e^{ik\cdot X(\tau,0)}\epsilon\cdot \partial_{\tau}X(\tau,0).
\end{equation}

From here on we set $Q=1$. In terms of the complex variable $z=e^{i\tau}$ this becomes

\begin{align}
\label{clasmultipoles}
\nonumber\mathcal{A}^{\text{cl}}=& \frac{1}{2\alpha'} \sum_{j=0}^{\infty}\left(-\sqrt{2\alpha'}\right)^j\frac{k_{a_1}\dots k_{a_j}}{j!}\oint \frac{dz}{2\pi i z}\Big(2\alpha'\epsilon\cdot {p}+\sqrt{2\alpha'}\sum_{n\neq 0}\epsilon\cdot\alpha_nz^{-n}\Big)\\&\times\Big(\sum_{m_1\dots m_j\neq 0}\frac{\alpha_{m_1}^{a_1}}{m_1}\dots \frac{\alpha_{m_j}^{a_j}}{m_j}z^{-m_1+\dots-m_j}\Big).
\end{align}

Performing the $z$ integral yields
\begin{equation}
\label{eq:classA3gen}
     \mathcal{A}^{\text{cl}}=\frac{1}{2\alpha'} \sum_{j=0}^\infty (-\sqrt{2\alpha'})^j\bigg[(2\alpha')(\epsilon\cdot {p})k_{a_1}\!\dots k_{a_j}T_{(j)}^{a_1\dots a_j}+\sqrt{2\alpha'}\epsilon_{a}k_{a_1}\!\dots k_{a_j}S_{(j)}^{a a_1\dots a_j}\bigg],
\end{equation}
where $T_{(0)} := 1$ and
\begin{align}
\label{eq:TandStensors}
    & T_{(j)}^{a_1\dots a_j}:=\frac{1}{j!}\sum_{m_1\dots m_j\neq 0} \frac{\alpha_{m_1}^{a_1}}{m_1}\dots  \frac{\alpha_{m_j}^{a_j}}{m_j}\delta_{\sum_{i=1}^jm_i,0},\\
    & S_{(j)}^{a a_1\dots a_j}:=\frac{1}{j!}\sum_{n,m_1,\dots m_j\neq 0}\alpha^{a}_n\frac{\alpha_{m_1}^{a_1}}{m_1}\dots  \frac{\alpha_{m_j}^{a_j}}{m_j}\delta_{n+\sum_{i=1}^jm_i,0}.    
\end{align}
Note that the above definition implies $S_{(0)}^a = T_{(1)}^a = 0$, and that
\begin{equation}
\label{eq:Sclass}
    S_{(1)}^{ab} := i S^{ab} =  \sum_{n>0}\frac{\alpha_{-n}^{a}\alpha_{n}^{b}-\alpha_{-n}^{a}\alpha_{n}^{b}}{n} ,
\end{equation}
recovering the usual definition of the classical spin tensor $S^{ab}$ for the string solution, obeying the covariant spin supplementary condition $p_aS^{ab} = 0$ due to the proper time gauge choice.

A few comments are now in order. In the previous sections we found that the infinite-spin amplitudes $\mathcal{A}^{(\infty)}$ are a function of the vectors $\epsilon^a$, $p^a$, $k^a$ and the spin tensor $S^{ab}$, for example in \eqn{eq:infAtworow}. On the other hand, the amplitude $\mathcal{A}^{\text{cl}}$ depends on the infinite set of tensors $T_{(j)}$ and $S_{(j)}$. This means that there can be a matching between the two, $\mathcal{A}^{(\infty)} = \mathcal{A}^{\text{cl}}$, only if we can find a special choice of the classical oscillators $\alpha_n^a$ such that all tensors $T_{(j)}$ and $S_{(j)}$ can be rewritten in terms of the spin tensor $S^{ab}$.

\subsection{Leading Regge warm-up}
\label{sect5.2}
Let us consider the following string solution,
\begin{align}
\label{eq:classsolLR}
\begin{split}
 & X^0 =2\alpha' m \tau , \\
 & X_1 = 2\sqrt{2\alpha'}\, b \sin \tau  \cos \sigma ,  \\
 & X_2 = -2\sqrt{2\alpha'}\, b \cos \tau  \cos \sigma ,  \\
 & X_3 = 0 ,
\end{split}
\end{align}
where $m^2 = 2b^2/\alpha'$ is required to satisfy the Virasoro constraints. We can easily extract the Fourier modes $\alpha_n$, which are $\alpha_{n>1} = 0$ and
\begin{equation}
\label{eq:classalpha1}
    \alpha_1^a = b(0,1,i,0) ,
\end{equation}
and we can obtain all other modes using the relation $\alpha_{-n} = (\alpha_n)^*$.
Note that this choice of $\alpha_n$ leads to a significant simplification of the $T_{(j)}$ and $S_{(j)}$ tensors defined in \eqn{eq:TandStensors}. First, we have $S_{(2j)} = T_{(2j+1)} = 0$ for all $j \in \mathbb{N}$. Moreover, we have
\begin{align}
\label{eq:STtoSpinLR}
\begin{split}
    k_{a_1}\cdots k_{a_{2j}}T_{(2j)}^{a_1\dots a_{2j}} &= \frac{(-k\cdot\alpha_{-1} k\cdot\alpha_{1})^j}{(j!)^2} = \frac{(k\cdot a\cdot a\cdot k)^j}{2\alpha' (j!)^2} , \\
\varepsilon_{a}k_{a_1}\dots k_{a_{2j+1}}S_{(2j+1)}^{a a_1\dots a_{2j+1}} &= \frac{(-k\cdot\alpha_{-1} k\cdot\alpha_{1})^j (f^{a b} \alpha_{1 a}\alpha_{-1 b})}{j!(j+1)!} \\&= \frac{-i m (f^{a b} a_{a b})(k\cdot a\cdot a\cdot k)^j}{2(2\alpha') j! (j+1)!} ,
\end{split}
\end{align}
where $f_{ab}=k_{[a}\epsilon_{b]}$, $a^{ab} := m^{-1} S^{ab}$ and where we used $k\cdot\alpha_{-1} k\cdot\alpha_{1} = -k\cdot S\cdot S\cdot k/(2\alpha_{-1}\cdot\alpha_1)$ from \eqn{eq:Sclass}, where $\alpha_{-1}\cdot\alpha_1 = \alpha'm^2$  from \eqn{eq:classalpha1}. Note that in this case all the $T_{(j)}$ and $S_{(j)}$ tensors have been rewritten in terms of the spin $S^{ab}$. Moreover, from \eqn{eq:classA3gen} we get
\begin{align}
\label{eq:classA3LR}
\mathcal{A}^{\text{cl}}_{{\mathrm{LR}}}=\epsilon\cdot p I_0(2\sqrt{k\cdot a\cdot a\cdot k})+i m\frac{f^{ab}a_{ab}}{2\sqrt{k\cdot a\cdot a\cdot k}}I_1(2\sqrt{k\cdot a\cdot a\cdot k}),
\end{align}
agreeing with the infinite-spin limit result in \eqn{eq:infspinA3LR}, i.e. $\mathcal{A}^{\text{cl}}_{\mathrm{LR}}=\mathcal{A}^{(\infty)}_{\mathrm{LR}}$, provided that $p\equiv p_1$.

\subsection{Two-mode solution}
\label{sect5.3}

Now let us proceed to the case of two harmonics. The most general string solution has form
\begin{align}
\label{eq:twomodeX}
\nonumber  X^a =& 2\alpha' p^a \tau + i\sqrt{2\alpha'} \left( \alpha_1^a e^{-i\tau} \cos{\sigma} - \alpha_{-1}^a e^{i\tau} \cos{\sigma} + \frac{1}{2}\alpha_2^a e^{-2i\tau} \cos{2\sigma} \right. \\& \left. - \frac{1}{2}\alpha_{-2}^a e^{2i\tau} \cos{2\sigma}  \right) .
\end{align}
The Virasoro constraints $\dot{X}\cdot X' = (\dot{X})^2+(X')^2 = 0$ yield the mass-shell constraint  $\alpha' m^2 = |\alpha_1|^2+|\alpha_2|^2$ as well as
\begin{equation}
\label{eq:twomodeVC}
    \alpha_1^2 = \alpha_2^2 = \alpha_1 \cdot \alpha_2 = \alpha_{-1} \cdot \alpha_2 = 0,
\end{equation}
and $\alpha_{-1}^2 = \alpha_{-2}^2 = \alpha_{-1} \cdot \alpha_{-2} = \alpha_{1} \cdot \alpha_{-2} = 0$ by complex conjugation. Recall that we are working in the centre-of-mass frame and in the proper time gauge in \eqn{BB4}, meaning $p\cdot\alpha_i = 0$. 

We can now look for a solution to \eqn{eq:twomodeVC}. Let us start from four dimensions. We can write
\begin{equation}
\alpha_1^a = a_1^a+ib_1^a~~~;~~~\alpha_2^a= a_2^a+i b_2^a
    \label{}
\end{equation}
The following constraints must be imposed,
\begin{align}
&\alpha_1^2=0 \Longrightarrow a_1^2=b_1^2 ~~;~~a_1\cdot b_1=0~~~;~~
\alpha_2^2=0 \Longrightarrow a_2^2=b_2^2~~;~~a_2\cdot b_2=0 \nonumber \\
& \alpha_1\cdot \alpha_2 = \alpha_{-1}\cdot \alpha_2 = 0 \Longrightarrow a_1\cdot a_2= b_1\cdot b_2=a_1\cdot b_2=a_2\cdot b_1=0
    \label{}
\end{align}
These are 8 conditions on 4 real three-dimensional vectors, and they require all such vectors to be orthogonal to each other, but this is impossible in a three-dimensional space. For example, we can put $a_1, b_1$ in a plane orthogonal to each other.
Then we can put either $a_2$ or $b_2$ along a third axis orthogonal to the previous plane. In three space dimensions there are no more directions to put the remaining vector such that it is orthogonal to the other three. But this is possible in four or more spatial dimensions.

In five spacetime dimensions and higher, we can find nontrivial solutions. One example is the following,
\begin{equation}
\label{eq:twomodealphas}
    \alpha_1^a = b(0,1,i\eta,1,-i\eta) \quad , \qquad \alpha_2^a = \sqrt{2} c(0,1,i\varepsilon,-1,i\varepsilon) ,
\end{equation}
 leading to the following classical string solution $X^a$,
\begin{align}
\label{eq:twomodesol}
\begin{split}
    &X^0 = 2 m \alpha' \tau ,\\
    &X^1 = 2\sqrt{2\alpha'} \left( b \sin{\tau} \cos{\sigma} + \frac{c}{\sqrt{2}} \sin{2\tau} \cos{2\sigma} \right) ,\\
    &X^2 = 2\sqrt{2\alpha'} \left( -\eta b \cos{\tau} \cos{\sigma} - \frac{\varepsilon c}{\sqrt{2}} \cos{2\tau} \cos{2\sigma} \right) ,\\
    &X^3 = 2\sqrt{2\alpha'} \left( b \sin{\tau} \cos{\sigma} - \frac{c}{\sqrt{2}} \sin{2\tau} \cos{2\sigma} \right),\\
    &X^4 = 2\sqrt{2\alpha'} \left( \eta b \cos{\tau} \cos{\sigma} - \frac{\varepsilon c}{\sqrt{2}} \cos{2\tau} \cos{2\sigma} \right) ,
\end{split}
\end{align}
where $b, c$ are free parameters and $\eta, \varepsilon = \pm 1$. This solves the Virasoro constraints provided we have $\alpha' m^2 = 4(b^2 + 2c^2)$. 

\begin{figure}[h!]
\centering
\includegraphics[scale=0.53]{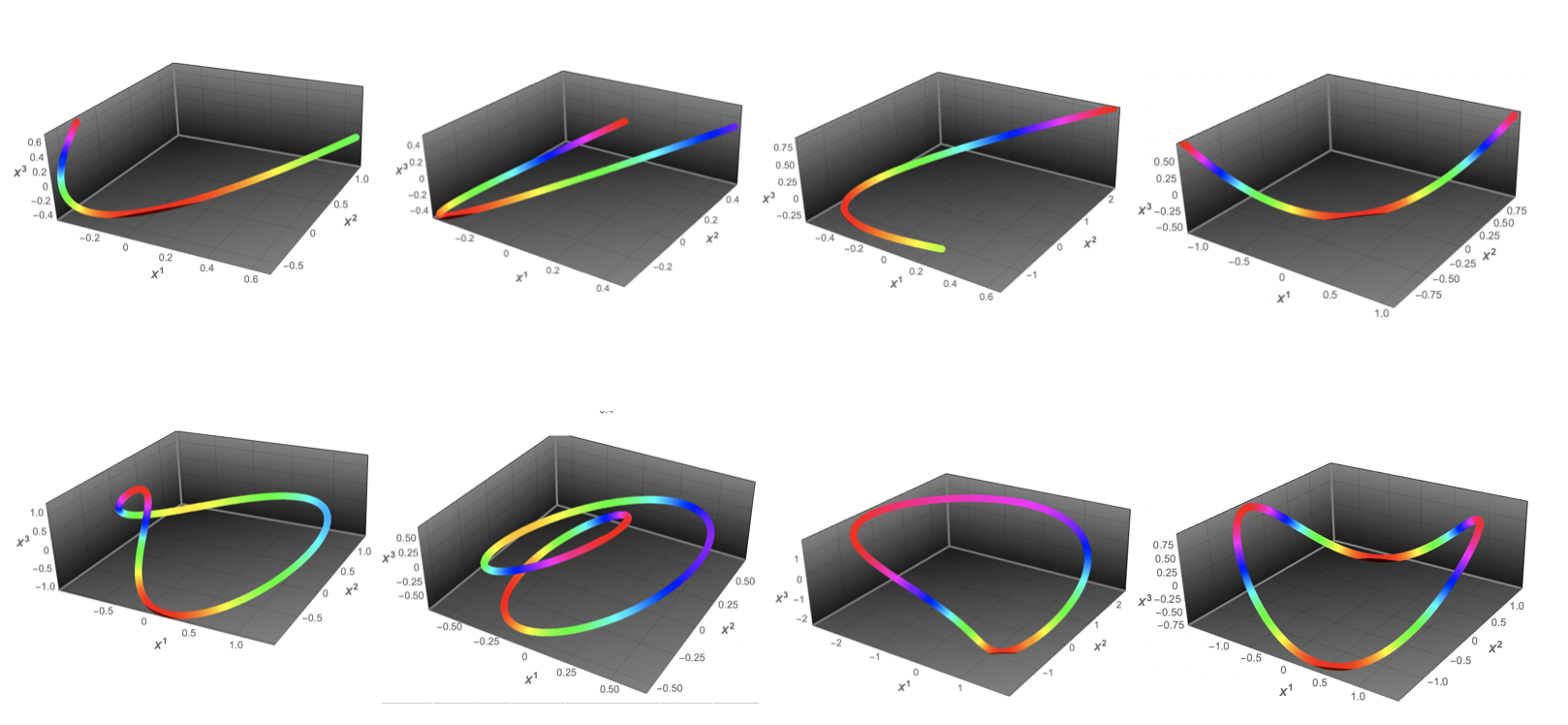}
\caption{In the first line from above there are three representative 3D profiles of the string solution at $\tau=0$. In the second line there are three representative profiles of the motion of the end point of the string at $\sigma=\pi$.   }
\label{classProf2harmEx}
\end{figure}

We also provide a generalisation of this solution in terms of six real parameters $\{r_1,s_1,s_2,\theta_1,\phi_1,\phi_2\}$ and four integer phases $\{\nu,\sigma,\rho,\delta\}$:
$$
\alpha_1^a = \left(0,(-)^\nu r_1 e^{i\theta_1},(-)^\sigma ir_1 e^{i\theta_1},(-)^{\nu+1}{r_1 s_1\over s_2}e^{-i\phi_1+i\phi_2+i\theta_1},(-)^{\sigma+1} i {r_1 s_1\over s_2}e^{-i\phi_1+i\phi_2+i\theta_1}\right)
$$
\be
\label{genSol}
 \alpha_2^a =\Big(0,(-)^{\rho}s_1 e^{i\phi_1},(-)^{\delta}i s_1 e^{i\phi_1},(-)^{\rho} s_2 e^{i\phi_2},(-)^{\delta}i s_2 e^{i\phi_2}\Big)
\ee
For illustration, in Fig.~\ref{classProf2harmEx} we report the 3D section and the motion of the end-point of the general string solution, for various random choices of the parameters.

Having found a classical solution, the next step is to derive the electromagnetic current. This is given by \eqn{eq:classA3gen}, and we can list a few of the tensors $T_{(j)}$ and $S_{(j)}$ below. Since $T_{(0)} = 1$ and $S_{(0)}^a = T_{(1)}^a = 0$, the first nontrivial tensor structure is
\begin{equation}
    \epsilon_a k_b S_{(1)}^{ab} = -\frac{i}{2}f_{ab} S^{ab} = f_{ab}\alpha_1^a\alpha_{-1}^b + \frac{1}{2}f_{ab}\alpha_2^a\alpha_{-2}^b.
\end{equation}
This is the usual spin-dipole contribution to the amplitude. The next contribution is $2\alpha' \epsilon\cdot p \, k_a k_b T_{(2)}^{ab} + \sqrt{2\alpha'} \epsilon_a k_b k_c S_{(2)}^{abc}$, where
\begin{align}
\label{eq:twomodequadrupole}
\begin{split}
    k_a k_b T_{(2)}^{ab} =& -k\cdot\alpha_{1}k\cdot\alpha_{-1}-\frac{1}{4}k\cdot\alpha_{2}k\cdot\alpha_{-2} , \\
    \epsilon_a k_b k_c S_{(2)}^{abc} =& \, \frac{1}{2} \left( (k\cdot\alpha_{1})^2\epsilon\cdot\alpha_{-2} + (k\cdot\alpha_{-1})^2\epsilon\cdot\alpha_{2} -k\cdot\alpha_{2} k\cdot\alpha_{-1} \epsilon\cdot\alpha_{-1} \right.\\
    &\left.\, - k\cdot\alpha_{-2} k\cdot\alpha_{1} \epsilon\cdot\alpha_{1}\right) .
\end{split} 
\end{align}

Note that, in the spin-multipole expansion of a classical amplitude, the term proportional to $k^a k^b$ is the quadrupole term, containing two powers of the spin tensor $S^{ab}$. Since $\epsilon\cdot k = p\cdot k=0$ and $p_a S^{ab} = 0$, the only allowed Lorentz invariant structure is $k\cdot S\cdot S\cdot k$, or equivalently $k\cdot a\cdot a\cdot k$, as we can see explicitly in results such as \eqn{eq:infAtworow} and \eqn{eq:classA3LR}. In this case, we have
\begin{equation}
\label{eq:twomodekSSk}
    k\cdot a\cdot a\cdot k = \frac{1}{m^2}\left(-2|\alpha_1|^2 k\cdot\alpha_{-1}k\cdot\alpha_{1}-\frac{1}{2}|\alpha_2|^2 k\cdot\alpha_{-2}k\cdot\alpha_{2}\right) ,
\end{equation}
using the constraints in \eqn{eq:twomodeVC}. If the two-mode solution considered here admits a standard spin-multipole expansion, then we should have
\begin{equation}
\label{eq:STtoSpinQuad}
    2\alpha' \epsilon\cdot p \, k_a k_b T_{(2)}^{ab} + \sqrt{2\alpha'} \epsilon_a k_b k_c S_{(2)}^{abc} = C_2 \, \epsilon\cdot p \, k\cdot a\cdot a\cdot k ,
\end{equation}
where $C_2$ is a constant proportional to the spin-quadrupole coefficient.

However, for a generic solution~\eqref{eq:twomodeX} satisfying the constraints~\eqref{eq:twomodeVC}, and for generic $k^a$ and $\epsilon^a$ compatible with three-point kinematics, it is clear that \eqn{eq:STtoSpinQuad} will not be satisfied. In particular, $S^{ab}$ contains an equal number of $\alpha_{n}$ and $\alpha_{-n}$ for each $n$ and therefore it can never give rise to the terms in $\epsilon_a k_b k_c S_{(2)}^{abc}$ in \eqn{eq:twomodequadrupole}. This is a clear difference to the leading Regge case, where we found \eqn{eq:STtoSpinLR} relating all $S_{(j)}$ and $T_{(j)}$ tensors to $S^{ab}$. This means that the classical current from the two-harmonics classical string solution~\eqref{eq:twomodeX} cannot agree with the infinite-spin amplitude~\eqref{eq:infAtworow} from the tilted trajectory. More in general, the currents of generic string solutions do not admit a spin-multipole expansion in terms of the spin tensor $S^{ab}$ and therefore are fundamentally different from the amplitudes obtained from limits along string trajectories.~\footnote{Here we are assuming that infinite-spin limits of string amplitudes always produce a spin-multipole expansion. This is true in all the cases we considered but it requires further investigation.}

Note that, in the discussion above, we assumed that $\alpha_{i}^a$ are arbitrary solutions of the Virasoro constraints, in analogy to the leading Regge case in \eqn{eq:STtoSpinLR}. However, one might wonder if there exists a special choice of $\alpha_{i}^a$ which does match the infinite-spin limit of a quantum amplitude. 
We can test this explicitly using the solution~\eqref{eq:twomodesol}. We can choose
\begin{equation}
    k^a = \omega(0,i,0,-1,0) \quad , \qquad \epsilon^a = \frac{1}{\sqrt{2}}(1,0,-1,0,0) .
\end{equation}
Using $p = (m,\vec{0})$ as usual, this choice satisfies three-point kinematics $p\cdot k =0$ and $k^2 = \epsilon\cdot k = \epsilon^2 = 0$. Note that the momenta have to be complex since this kinematics does not admit a real solution. Then, using the values of $\alpha_i$ in \eqn{eq:twomodealphas} we find that the quadratic-in-$k$ term of the classical amplitude~\eqref{eq:classA3gen} is
\begin{equation}
    2\alpha' \ep\cdot p \, k_a k_b T_{(2)}^{ab} + \sqrt{2\alpha'} \ep_a k_b k_c S_{(2)}^{abc} = -\frac{4i\sqrt{2}(2b^2-c^2)(b^2+2c^2)\omega^2}{m} .
\end{equation}
If this is to be interpreted as a spin-quadrupole term, it should be proportional to $k\cdot a\cdot a \cdot k$. We find
\begin{equation}
    \epsilon\cdot p \,\, k\cdot a\cdot a\cdot k = -\frac{8i\sqrt{2}(b^4-c^4)\omega^2}{m} .
\end{equation}
This would seem to fix $C_2 = (2b^2-c^2)(b^2+2c^2)/(2b^4-2c^4)$ in \eqn{eq:STtoSpinQuad}. However, if this is true, it should hold independently of the chosen values for $k$ and $\epsilon$. Hence, we can check by choosing $k = \omega(0,i,-1,0,0)$ and $\epsilon = \frac{1}{\sqrt{2}}(1,0,0,-1,0)$, which also satisfy three-point kinematics and the on-shell conditions. In this case we find
\begin{equation}
    2\alpha' \epsilon\cdot p \, k_a k_b T_{(2)}^{ab} + \sqrt{2\alpha'} \epsilon_a k_b k_c S_{(2)}^{abc} = -2\sqrt{2\alpha'} \omega^2 b^2 c ,
\end{equation}
but we have $k\cdot a \cdot a \cdot k = 0$. For generic values of $b, c$ this is clearly a contradiction and it confirms the previous argument that the two-mode solution does not admit a spin-multipole expansion. The only choice that may resolve this contradiction is $b = c$, leading to $C_2 \to \infty$ for both choices of $k^a$ and $\epsilon^a$. However, we still need to check that all other tensors $S_{(j)}$ and $T_{(j)}$ reduce to spin tensors for this special choice. We find this is not the case, and we omit details for conciseness. This analysis can also be extended to the more general solution~\eqref{genSol}, but even then we observed a similar behaviour.

\section{String coherent states}
\label{sect6}

Aside from $s\to\infty$ limits of string trajectories, string coherent states are  natural objects to consider when trying to reproduce classical observables. A coherent state is defined as
\begin{equation}
    \ket{b_1,\dots,b_m} = e^{-\frac{1}{2}\sum_{i>0}^m|b_m|^2} e^{\sum_{n>0}^mb_n \cdot a_n^\dagger} \ket{0} ,
\end{equation}
where $b_n$ are complex vectors subject to the Virasoro constraints, as discussed below. Since $[a_n^a,a_n^{\dagger b}] = \eta^{a b}$, we have
\begin{equation}
\label{eq:coheigen}
    a_n^a \ket{b_1,\dots,b_m} = b_n^a \ket{b_1,\dots,b_m} ,
\end{equation}
where $b_{n>m}^a := 0$. Namely, $\ket{b_1,\dots,b_m}$ is an eigenvector of all annihilation operators. Having defined the state $\ket{b_1,\dots,b_m}$, we can use it to compute the expectation value of any observable $\hat{\cal O}$ as \begin{align}
\langle \hat{\mathcal{O}} \rangle_m :=\bra{b_1,\dots,b_m} \hat{\mathcal{O}}\ket{b_1,\dots,b_m}.
\end{align}
We can also impose the Virasoro constraints on coherent states, in the following fashion,~\footnote{Unlike the quantum case where it is sufficient to impose only the physical conditions involving $L_1$ and $L_2$ because the others are automatically satisfied following from the Virasoro algebra, in the classical case we have to impose all the conditions in \eqref{BB5}. }
\begin{equation}
\langle   L_n  \rangle_m =0 \quad,\quad n\geq 0 .
    \label{BB5}
\end{equation}
Note that this is different from the way Virasoro constraints are imposed on quantum states, e.g. \eqn{eq:virasoro}. Also, given the definitions $L_n = (1/2)\sum_{k\in\mathbb{Z}} \hat\alpha_{k}\cdot\hat\alpha_{n-k}$, $a_n^\dagger = \hat\alpha_{-n}/\sqrt{n}$ and $a_n = \hat\alpha_{n}/\sqrt{n}$ for $n>0$, using \eqn{eq:coheigen} we get
\begin{equation}
   \langle   L_n  \rangle_m = \frac{1}{2} \sum_{k\in\mathbb{Z}} \sqrt{k}\sqrt{n-k}\, b_k\cdot b_{n-k} ,
\end{equation}
meaning that the Virasoro constraints for coherent states are identical to those for classical string solutions in \eqn{eq:virasoro2}, upon the identification $\alpha_n^a = \sqrt{n} b_n^a$.
For the sake of simplicity we work in the center-of-mass frame $p = (m,\vec{0})$ where the coefficients $b_n^a$ obey $b_n^0 = 0$. The first observable we consider is the string position operator, focussing on the oscillator part,
\begin{align}
& {\hat{X}}^a = i \sqrt{2\alpha'} \sum_{n=1}^\infty \frac{1}{\sqrt{n}}
\bigg( a_{n}^a e^{-in\tau} - a_{n}^{\dagger a} e^{in\tau}\bigg) \cos (n\sigma) \,\,.
\label{BB7}
\end{align}
From \eqn{eq:coheigen} we get
\begin{equation}
\label{eq:cohX}
  \langle \hat{X}^a\rangle_m= i \sqrt{2\alpha'} \sum_{n=1}^m\frac{1}{\sqrt{n}}
\bigg( b_{n}^a e^{-in\tau} - b_{n}^{* a} e^{in\tau}\bigg) \cos (n\sigma) .
\end{equation}
Once again, this is equivalent to the classical result~\eqref{eq:genclassX} upon identifying $\alpha_n^a = \sqrt{n} b_n^a$. Therefore there is a one-to-one correspondence between classical string solutions and string coherent states, at least at the level of the free solution $X^a$. Note that the zero-mode part agrees trivially due to the relation $\hat{p}^a\ket{b} = p^a\ket{b}$.

The next observable we can consider is the vertex operator for a massless vector, namely
\begin{equation}
    \hat{\mathcal{A}}=\frac{1}{4\pi\alpha'}\int_0^{2\pi}d\tau \, \epsilon\cdot\partial_{\tau}\hat{X}(\tau,0) e^{ik\cdot \hat{X}(\tau,0)} .
\end{equation}
Note that in this case the quantity $\langle \hat{\mathcal{A}}\rangle_m$ is just the three-point amplitude between two coherent states and the massless vector, similarly to the quantum case discussed in \eqn{3ptLR} and \eqn{3pt2HaZet}. We can rewrite the integrand in terms of the ladder operators as
\begin{align}
\begin{split}
    \epsilon\cdot\partial_{\tau}\hat{X}(\tau,0) =& \, 2\alpha'\epsilon\cdot p+\sqrt{2\alpha'} \sum_{n=1}^\infty \sqrt{n}\big( \epsilon\cdot a_{n} e^{-in\tau} + \epsilon\cdot a_{n}^{\dagger} e^{in\tau} \big) , \\
    ik\cdot \hat{X}(\tau,0) =& - \sqrt{2\alpha'} \sum_{n=1}^\infty \frac{k\cdot a_{n} e^{-in\tau} - k\cdot a_{n}^{\dagger} e^{in\tau}}{\sqrt{n}} .
\end{split}
\end{align}
using $p\cdot k = 0$. When evaluating $\langle\hat{\mathcal{A}}\rangle_m$, in principle we have to expand the exponential and be careful to the commutation relations between the operators $\epsilon\cdot a_n, \epsilon\cdot a_n^\dagger, k\cdot a_n$ and $k\cdot a_n^\dagger$. However, since $k^2 =  \epsilon\cdot k = 0$, all commutators vanish and we can trivially apply \eqn{eq:coheigen} to get
\begin{equation}
\label{eq:cohA3}
   \langle\hat{\mathcal{A}}\rangle_m = \frac{1}{4\pi\alpha'}\int_0^{2\pi}d\tau e^{ik\cdot \langle \hat{X}(\tau_,0)\rangle_m} \epsilon\cdot\partial_{\tau}\langle \hat{X}(\tau,0)\rangle_m.
\end{equation}
Since we found in \eqn{eq:cohX} that $\langle \hat{X}^a\rangle_m$ is equivalent to the classical solution $X^a$, we conclude that the coherent-state amplitude~\eqref{eq:cohA3} is equal to the classical amplitude~\eqref{eq:classA3} upon identifying $\alpha_n^a = \sqrt{n} b_n^a$, i.e. \begin{align}
    \langle\hat{\mathcal{A}}\rangle_m=\mathcal{A}^{\mathrm{cl}}.
\end{align} This is an important result, since it means that coherent states and classical solutions are in one-to-one correspondence not only at the level of the free solution $X^a$, but also at the level of the three-point amplitude.

We are now ready to draw some conclusions about the correspondence between classical solutions and string states:
\begin{itemize}
    \item We have shown that the coherent-state amplitude and the amplitude defined from the classical solution will always be identical;
    \item We have seen that the amplitude for a single-mode classical solution~\eqref{eq:classsolLR} agrees with the amplitude for the leading-Regge state $(\ep\cdot a_1^\dagger)^s\ket{0}$ for $s\to\infty$. Both also agree with the amplitude for a single-mode coherent-state $\exp(b_1\cdot a_1^\dagger)\ket{0}$;\footnote{Note that the coherent-state amplitude is also the generating function~\eqref{eq:GfinalLR} for the state $(\ep\cdot a_1^\dagger)^s\ket{0}$. Indeed, we can see that it has the same functional form as the $s\to\infty$ amplitude~\eqref{eq:infspinA3LRbesselJ}.}
    \item In the case of two harmonics, the classical amplitude for the string solution~\eqref{eq:twomodeX} is guaranteed to agree with the amplitude in terms of the coherent state $ \exp(b_1\cdot a_1^\dagger + b_2\cdot a_2^\dagger)\ket{0}$. However, these results are different from the $s\to\infty$ limit of the tilted trajectory states $\big(\zeta_{ab}a_1^{\dagger a}a_2^{\dagger b}\big)^s\ket{0}$. One possible explanation for this is that in this case the coherent state contains those states as well as many other string states, whereas in the one-harmonic case it only contains leading Regge states. 
\end{itemize}
\subsection{Classical observables from coherent states}
\label{sect6.1}

In this Section we explicitly use the coherent states introduced above to recover the classical string solutions discussed in section~\ref{sect5} as well as the classical currents.

The simplest classical solutions are those where the string moves in a plane. In this case, following the general approach described before, we  start from the one-oscillator coherent state $|b_1 \rangle \propto e^{b_1\cdot a_{1}^\dagger}|0\rangle $ and impose \eqref{BB5}. In the center-of-mass frame, where $b_1^a=(0,b_1^i)$, the only nontrivial constraints come from $\langle L_2\rangle_1 = 0$ and require $b_1^2=(b^*_1)^2 = 0$ so that we can choose $$b_1^a = (0,b,ib,0),$$ and hence
\begin{equation}
 |b_1 \rangle = e^{-b^2}e^{b(a_{n}^{1\dagger} + i a_{n}^{2\dagger})}|0\rangle,
    \label{BB9}
\end{equation}
where $b$ is an arbitrary real parameter.

The classical solution we get is again \eqn{eq:classsolLR} and its mass and angular momentum are
\begin{align}
\langle \hat{S}^{ab}\rangle_1=-ib_1^{*[a} b_1^{b]},\qquad \alpha'm^2=|b_1|^2=2b,
\end{align}
so that, by denoting the only non-vanishing component of $\langle \hat{S}^{ab}\rangle_1$ as $\langle \hat{S}^{12}\rangle_1:= S=2b^2$, we get 
\begin{align}
\alpha' m^2 = S,
\end{align}
which is indeed the leading-Regge trajectory. The expectation value of the massless-vector vertex operator \eqref{eq:cohA3} is the same as \eqref{eq:GfinalLR}, where we replace $(c_1\varepsilon_1^{a},c_2\varepsilon_2^a)$ by $(b_1^{*a},b_1^a)$ and thus it reads
\begin{align}
\langle\hat{\mathcal{A}}\rangle_1=\bigg(-\epsilon\cdot p J_0(2\sqrt{2\alpha'|b_1\cdot k|^2})+f_{ab}b_1^{*a}b_1^b\frac{J_1(2\sqrt{2\alpha'|b_1\cdot k|^2)}}{\sqrt{2\alpha'|b_1\cdot k|^2)}}\bigg).
\end{align}
It is then immediate to show that this can be rewritten as 
\begin{align}
\langle\hat{\mathcal{A}}\rangle_1=\bigg(\epsilon\cdot p \,I_0(2\sqrt{k{\cdot}\langle a\rangle_1 {\cdot}  \langle a\rangle_1{\cdot} k})+im\frac{f^{ab}\langle a^{ab}\rangle_1}{2\sqrt{k{\cdot} \langle a\rangle_1{\cdot} \langle a\rangle_1{\cdot} k}}I_1(2\sqrt{k{\cdot} \langle a\rangle_1{\cdot}\langle a\rangle_1{\cdot} k})\bigg),
\end{align}
with $\langle a^{ab}\rangle_1=\langle \hat{S}^{ab}\rangle_1 m^{-1}$. Thus we see explicitly that $\langle\hat{\mathcal{A}}\rangle_1$ is equal to the classical electromagnetic current $\mathcal{A}^{\mathrm{cl}}_{\mathrm{LR}}$ computed in \eqref{eq:classA3LR}, as discussed in the previous Section.

Let us consider now the two harmonics case, where the coherent state  has the following form:
\begin{equation}
|b_1,b_2\rangle =e^{-\frac{1}{2}(|b_1|^2+|b_2|^2)} e^{b_1\cdot a_1^\dagger} e^{b_2 \cdot a_2^\dagger}|0\rangle,
\end{equation}
By imposing the condition in \eqref{BB5} one gets:
\begin{equation}
b_1^*\cdot b_2=b_1^2=b_1\cdot b_2 = b_2^2=0,
    \label{BB13}
\end{equation}
The first condition is obtained from $L_1$, the second  from $L_2$,  the third  from $L_3$ and the fourth from $L_4$. They are $8$ real conditions imposed on four real $D-1$-dimensional quantities.  For $D\leq 4$ there is no solution. The first solution appears for $D=5$ and it is given, for example, by 
\begin{align}
&b_1^a= b(0,1, i\eta, 1, -i \eta) ~~~ ; ~~~ 
b_2^a= c (0,1, i\varepsilon, -1, i\varepsilon),
    \label{BB14}
\end{align}
where $\eta^2 = \varepsilon^2=1$,
leading to the following coherent state,
\begin{align}
\label{2hcs}
& |b_1,b_2\rangle=  e^{b ( a_{1}^{1\dagger} + i \eta a_{1}^{2\dagger} +a_{1}^{3\dagger}-i\eta a_{1}^{4\dagger})} \quad  e^{c(a_{2}^{1\dagger} +i\varepsilon a_{2}^{2\dagger} - a_{2}^{3\dagger}+i\varepsilon a_{2}^{4\dagger})} |0\rangle.
\end{align}
Evaluating \eqref{eq:cohX} with the two-harmonic state in \eqref{2hcs} one gets again the classical string solution in \eqn{eq:twomodesol}. For the mass and angular momentum one gets:
\begin{equation}
\langle \hat{S}^{ab}\rangle_2=-i(b_1^{*[a}b_1^{b]}+b_2^{*[a}b_2^{b]}),\qquad\alpha' m^2 =4 (b^2+2c^2),
\end{equation}
yielding 
\begin{align}
\alpha'm^2=2\frac{b^2+2c^2}{c^2\varepsilon+b^2\eta}S,
\end{align}
which, for $b=c$ and $\varepsilon=\eta=1$, gives $\alpha'm^2=3S$, i.e. the classical limit of the quantum trajectory considered in Sect.~\ref{Sect:2HS}. However, even for this special choice of parameters, the $s\rightarrow\infty$ limit of the amplitude therein considered will not match the coherent state one, as previously discussed.

\section*{Conclusion and outlook}
\label{conclu}

This paper continues the approach of \cite{Cangemi:2022abk}
in trying to find the classical limit of the three-point amplitude involving two massive spinning particles and a graviton in the bosonic string corresponding to the Kerr black hole linearised energy-momentum tensor. For simplicity, we have restricted ourselves to the electromagnetic case that goes under the name of $\sqrt{\mathrm{Kerr}}$. 

Both in \cite{Cangemi:2022abk} and here one considers the infinite-spin limit along a chosen trajectory of massive states. However, since taking such limit along the leading Regge trajectory does not yield the expected Kerr behaviour, in this paper we consider another trajectory, forming an angle with respect to the leading one, that we have called tilted trajectory. 

This straight line contains physical states, involving only the oscillators $a_1$ and $a_2$,
that are described by Young diagrams having only two rows with length $s$ ($s=1,2 \dots \infty$) and the same number of columns as in \eqn{eq:youngtworow}. These are zero-depth states, according to the notation of \cite{Markou:2023ffh,Basile:2024uxn}, and this means that they are described by Young diagrams that appear for the first time in the low-energy spectrum.  
We perform the classical infinite-spin limit along this trajectory and, similarly to the leading-Regge case studied in Ref.~\cite{Cangemi:2022abk} and reviewed in this work, we get a sum of a few Bessel functions that do not reproduce the exponential behavior of $\sqrt{\mathrm{Kerr}}$, for which a sum of infinite Bessel functions is needed.

We then construct a classical string solution with two harmonics and we compute the electromagnetic current generated by the motion of a charge at one of the string endpoints and, unlike the case of the leading trajectory, we do not find agreement with the three-point amplitude previously computed in the $s \rightarrow \infty$ limit.

In the final part of this conclusive Section we want to discuss some facts that may help understand our results and inform future directions.

First of all, we notice that the states of the tilted trajectory are not the only physical states consisting of the oscillators $a_1$ and $a_2$. There are many other states of this kind, corresponding to Young diagrams with two rows, the first of length $s$ and the second of length $s-n$ for $s=1, 2,3 \dots \infty$ and $n=0, 1 , 2\dots s-1$. The case $n=0$ corresponds to the tilted trajectory. The others are additional states that may turn out to be relevant to get the precise matching between the infinite-spin limit of string amplitudes and the two-harmonic classical string solution.
Note that both these states and the classical string solution exist only for $D>4$, if the presence of compact dimensions is excluded. This coincidence is probably not accidental and it is compatible with the assumption that the classical solution provides a description of all these additional states. This may also explain why the coherent states with two harmonics, which contains all such states, provide instead the correct result. We plan to come back to these points hoping to clarify them in a future work.

However, it is likely that it is not 
possible to reproduce the $\sqrt{\mathrm{Kerr}}$ behaviour with only two harmonics. We remind here that, in order to get $\sqrt{\mathrm{Kerr}}$ behaviour, we need a sum involving an infinite number of Bessel functions,
\begin{equation}
\sum_{n=-\infty}^{+\infty} I_n (k\cdot a) = I_0(k\cdot a) + 2\sum_{n=0}^\infty I_n(k\cdot a) = e^{k\cdot a} ,
    \label{}
\end{equation}
in $D=4$. Since the leading Regge case yields $I_{n\leq 1}$ and the two-mode case yields $I_{n\leq 3}$, for $n \geq 0$, it is likely that a state that reproduces $\sqrt{\mathrm{Kerr}}$ contains infinitely-many modes.

We conclude stressing a few of our results. We constructed a new classical solution of the open string equations of motion and constraints and we computed novel three-point amplitudes for an infinite set of physical states with the first two harmonics. We studied the classical infinite-spin limit of these amplitudes and, in order to do so, we generalised the classical limit technology to states with mixed-symmetry Young diagrams in arbitrary dimensions.
Moreover, we showed that one obtains the same three-point amplitudes involving these states and a massless vector using both the DDF and the covariant formalism. We also provided a general procedure for constructing all possible classical solutions by means of
the string coherent states discussed in Sect.~\ref{sect6}, and we showed that coherent-state three-point amplitudes always reproduce the electromagnetic current of the corresponding solution.

\section*{Acknowledgments}
We would like to thank Massimo Bianchi, Bruno Bucciotti, Lucile Cangemi, Marco Chiodaroli, Simon Ekhammar, Roberto Emparan,  Alessandro Georgoudis, Henrik Johansson, Juan Maldacena, Chrysoula Markou, Rodolfo Russo,  Oliver Schlotterer, Marija Tomašević, Gabriele Veneziano for discussions and comments.
MF is supported by the European MSCA grant HORIZON-MSCA- 2022-PF-01-01 ``BlackHoleChaos'' No.101105116 and partially supported by the H.F.R.I call ``Basic research Financing (Horizontal support of all Sciences)'' under the National Recovery and Resilience Plan ``Greece 2.0'' funded by the European Union – NextGenerationEU (H.F.R.I. Project Number: 15384). The work of P.P. was supported by the Science and Technology Facilities Council (STFC) Consolidated Grants ST/T000686/1 “Amplitudes, Strings \& Duality” and ST/X00063X/1 “Amplitudes, Strings \& Duality”; no new data were generated or analysed during this study. One of us (PDV) thanks the Galileo Galilei Institute for Theoretical Physics for hospitality and the INFN for partial support during the completion of this work.

\begin{appendix}
\section{Generating functions of three-point amplitudes}
In this appendix we derive, using coherent-state techniques, the generating functions of the three-point amplitudes analysed in the paper.
\label{appA}
\subsection{Generating function for leading-Regge states}\label{GenffLR}
Let us start by considering the following correlation function
\begin{align}\label{CFLR}
{\cal C}_{\varepsilon_1,\varepsilon_2}(\beta_i):=\langle0|e^{\varepsilon_1\cdot a_1+\beta_1\cdot a_1}|e^{\varepsilon_2\cdot a_1^{\dagger 
}+\beta_2\cdot a_1^{\dagger}}|0\rangle ,
\end{align}
where 
$\beta^a_i$ are, in general, complex parameters. We will make use of the  coherent states completeness relation,
\begin{align}
\label{completeness}
\mathds{1}=\int \frac{d^{D}v_1}{\pi^D}\dots\int\frac{d^{D}v_n}{\pi^D}|v_1,\dots,v_n\rangle\langle v_1,\dots, v_n| \,e^{-|v_1|^2}\dots e^{-|v_n|^2},
\end{align}
where $|v_i\rangle$ are coherent states for the $i$-th oscillator,
\begin{align}
\label{eq:coherent}
|v_i\rangle=e^{-\frac{1}{2}|v_i|^2}e^{ v_i\cdot  a^{\dagger}_i}|0\rangle,\qquad v_i\in\mathbb{C}^D,\qquad |v_1,\dots,v_n\rangle=\otimes_{i=1}^n|v_i\rangle.
\end{align}
In \eqref{completeness}  $d^{D}v_i$ is the
standard measure over the complex plane of a $D$ dimensional vector. Inserting \eqref{completeness} with $n=1$ into \eqref{CFLR} we get 
\begin{align}
\label{ILR}
{\cal C}_{\varepsilon_1,\varepsilon_2}(\beta_i)=\int\frac{d^{D} v_1  }{\pi^D}e^{-|v_1|^2+(\varepsilon_1+\beta_1)\cdot v_1+(\varepsilon_2+\beta_2)\cdot \bar{v}_1}=e^{(\varepsilon_1+\beta_1)\cdot(\varepsilon_{2}+\beta_2)}.
\end{align}
The generating function in \eqref{LRgenFF}, which is
\be
\mathcal{A}^{(\mathrm{gen})}_{\,\mathrm{LR}}:=\langle p_{1}| e^{c_{1}\varepsilon_{1}{\cdot}a_{1}} \,\oint {dz\over 4\alpha'\pi i\,z}\, {\epsilon}{\cdot}i\partial \hat{X}\,e^{ik{\cdot}\hat{X}}(z)\, e^{c_{2}\varepsilon_{2}{\cdot}a^{\dagger}_{1}}|p_{2}\rangle\,,
\ee
can be written as
\be\label{GefLRapp}
\mathcal{A}^{(\mathrm{gen})}_{\,\mathrm{LR}}=\frac{1}{4\pi\alpha'}\int_0^{2\pi}d\tau\frac{\partial}{\partial\gamma}\langle p_1|e^{c_1\varepsilon_1\cdot a
_1}e^{\gamma\epsilon\cdot\partial_{\tau}\hat{X}(\tau)+ik\cdot \hat{X}(\tau)}e^{c_2\varepsilon_2\cdot a^{\dagger}
_1}|p_2\rangle\big|_{\gamma \mapsto 0}\,,
\ee
where we used the standard conformal mapping from the complex plane to the world-sheet cylinder, $z=e^{i\tau+\sigma}$, restricted to the boundary $\sigma=0$ for open string states. In \eqref{GefLRapp} we defined $\hat{X}(\tau):=\hat{X}(\sigma{=}0,\tau)$ and we exponentiated the massless vertex by introducing an auxiliary variable $\gamma$. The massless state is on-shell and $(\epsilon,k)$ satisfy $\epsilon\cdot k=\epsilon^2=k^2=0$. 
Using in \eqref{GefLRapp} the string mode expansion
\begin{equation}
\label{modexp}
    \hat{X}^a(\sigma,\tau) = 2\alpha'\hat{p}^a \tau + i\sqrt{2\alpha'} \sum_{n=1}^\infty \frac{1}{\sqrt{n}} \cos{n\sigma} (a_n^a e^{-i n\tau}-a_n^{\dagger a} e^{+i n\tau}) ,
\end{equation}
with $[a^a_n,a^{\dagger b}_m] = \eta^{ab} \delta_{n m}$, we get, from the zero modes, the standard momentum conserving delta function $\delta^{(D)}(p_1{+}k{+}p_2)$ and we find that 
\begin{align}
\label{eq:genfunctLRtac}
\mathcal{A}^{(\mathrm{gen})}_{\,\mathrm{LR}}=\frac{1}{4\pi\alpha'}\int_0^{2\pi}d\tau\frac{\partial}{\partial\gamma}e^{2\alpha'\gamma\epsilon\cdot p_2}{\cal C}_{c_1\varepsilon_1,c_2\varepsilon_2}(\beta_1(\tau),\beta_2(\tau))\big|_{\gamma\mapsto 0}\,,
\end{align}
where
\begin{align}
\label{ident1}
&\beta_1^{b}(\tau)=\sqrt{2\alpha'}(\gamma\epsilon^{b}-k^{b})e^{-i\tau},\qquad \beta_2^b(\tau)=\sqrt{2\alpha'}(\gamma\epsilon^{b}+k^{b})e^{i\tau}.
\end{align}
Notice that the derivative of the exponential factor in \eqref{eq:genfunctLRtac} gives the tachyon contribution to the amplitude.
Evaluating the $\tau$ integral gives the result in \eqref{eq:GfinalLR}.

\subsection{Generating function for next-to-leading Regge states}
\label{appNLR}
We start by computing the following matrix element
\begin{align}
\nonumber\mathcal{G}_{\varepsilon_1,\varepsilon_2}(\beta_i,\lambda_i):&=\zeta^1_{ab}\zeta^2_{cd}\langle 0|e^{\varepsilon_1\cdot a_1+\beta_1\cdot a_1+\lambda_1\cdot a_2}a_1^aa_2^b|e^{\varepsilon_2\cdot a_1^{\dagger}+\beta_2\cdot a^{\dagger}_1+\lambda_2\cdot a^{\dagger}_2}a^{\dagger d}_1a_2^{\dagger c}|0\rangle\\&=\zeta^1_{ab}\zeta^2_{cd}\mathcal{G}^{abcd}_{\varepsilon_1,\varepsilon_2}(\beta_i,\lambda_i),
\end{align}
with $\beta_i^a$ and $\lambda_i^a$ complex parameters and
\begin{align}
\mathcal{G}^{abcd}_{\varepsilon_1,\varepsilon_2}(\beta_i,\lambda_i):=\langle 0|e^{\varepsilon_1\cdot a_1+\beta_1\cdot a_1+\lambda_1\cdot a_2}a_1^aa_2^b|e^{\varepsilon_2\cdot a^{\dagger}_1+\beta_2\cdot a^{\dagger}_1+\lambda_2\cdot a^{\dagger}_2}a^{\dagger c}_1a_2^{\dagger d}|0\rangle.
\end{align}
Inserting the completeness relation \eqref{completeness} with $n=2$ we get
\begin{align}
  \nonumber\mathcal{G}^{abcd}_{\varepsilon_1,\varepsilon_2}(\beta_i,\lambda_i)&=\int\frac{d^Dv_1}{\pi^D}  \int\frac{d^Dv_2}{\pi^D}v_1^a\bar v_1^cv_2^b\bar v_2^d e^{-|v_1|^2-|v_2|^2+(\varepsilon_1+\beta_1)\cdot v_1+(\varepsilon_2+\beta_2)\cdot\bar v_1+\lambda_1\cdot v_2+\lambda_2\cdot\bar v_2}\\&=\frac{\partial}{\partial\beta_{1a}}\frac{\partial}{\partial\lambda_{1b}}\frac{\partial}{\partial\beta_{2c}}\frac{\partial}{\partial\lambda_{2d}}\tilde{\mathcal{G}}_{\varepsilon_1,\varepsilon_2}(\beta_i,\lambda_i),
\end{align}
with 
\begin{align}
\tilde{\mathcal{G}}_{\varepsilon_1,\varepsilon_2}(\beta_i,\lambda_i):=\int\frac{d^Dv_1}{\pi^D} \int\frac{d^Dv_2}{\pi^D}e^{-|v_1|^2-|v_2|^2+(\varepsilon_2+\beta_2)\cdot \bar{v}_1+(\varepsilon_1
+\beta_1)\cdot v_1+\lambda_2\cdot \bar{v}_2+\lambda_1\cdot v_2}.
\end{align}
Performing the Gaussian integrals we obtain
\begin{align}
\nonumber\mathcal{G}_{\varepsilon_1,\varepsilon_2}(\beta_i,\lambda_i)=\zeta^1_{ab}\zeta^2_{cd}[\eta^{bd}+\lambda_1^{d}\lambda_2^b][\eta^{ac}+(\varepsilon^c_1+\beta_1^{c})(\varepsilon^a_2+\beta_2^a)]e^{(\varepsilon_{2}+\beta_{2})\cdot(\varepsilon_1+\beta_1)+\lambda_1\cdot\lambda_2}.
\end{align}
The generating function~\eqref{defgenNLR} for the amplitude involving the states in \eqref{stateNLR} and a massless vector is
\begin{align}
\mathcal{A}^{(\mathrm{gen})}_{\mathrm{NLR}}:=\zeta^1_{ab}\zeta^2_{cd}\langle p
_1|e^{c_1\varepsilon_1\cdot a_1}a_1^aa_2^b\oint\frac{dz}{4\alpha'\pi i z}\epsilon\cdot i\partial \hat{X}\,e^{ik\cdot \hat{X}}(z)\,e^{c_2\varepsilon_2\cdot a^{\dagger}_1}a_1^{\dagger c}a_2^{\dagger d}|p_2\rangle,
\end{align}
and it can be recast as 
\begin{align}
\label{tauintnlr}
\nonumber \mathcal{A}^{(\mathrm{gen)}}_{\mathrm{NLR}}&=\frac{\zeta^1_{ab}\zeta^2_{cd}}{4\pi\alpha'}\int_0^{2\pi}d\tau \frac{\partial}{\partial\gamma} \langle p_1|e^{c_1\varepsilon_1\cdot a_1}a_1^aa_2^b\,e^{\gamma\epsilon\cdot \partial \hat{X}(\tau)+ik\cdot \hat{X}(\tau)}\,e^{c_2\varepsilon_2\cdot a^{\dagger}_1}a_1^{\dagger c}a_2^{\dagger d}|p_2\rangle\big|_{\gamma \mapsto 0}\\&=\frac{1}{4\pi\alpha'}\int_0^{2\pi}d\tau\frac{\partial}{\partial\gamma}e^{2\alpha'\gamma\epsilon\cdot p_2}\mathcal{G}_{c_1\varepsilon_1,c_2\varepsilon_2}(\beta_i(\tau),\lambda_i(\tau))\big|_{\gamma \mapsto 0},
\end{align}
where $\beta_i(\tau)$ are given in \eqref{ident1} and
\begin{align}
\label{relparameters}
&\lambda^b_1(\tau)= \sqrt{\alpha'} (2\gamma\epsilon^b-k^b) e^{-2i\tau},\qquad\lambda^b_2(\tau)= \sqrt{\alpha'} (2\gamma\epsilon^b+k^b) e^{2i\tau}.
\end{align}
Performing the $\tau$ integral in \eqref{tauintnlr}, gives \eqref{eq:genfNLR}.

\subsection{Generating function for tilted-trajectory states}\label{GenffTwH} 
In this appendix we repeat the procedure described above for the case of the state~\eqref{eq:tworowfactorepsilon}.
We start by computing the following quantity 
\be
 {\cal{I}}_{\zeta_1,\zeta_2}(\beta_i,\lambda_i):= \langle 0| e^{\zeta_1^{cd} a_{1c} a_{2d}+ \beta_1\cdot a_{1}  + \lambda_1\cdot a_{2}} | e^{\zeta_2^{cd} a^{\dagger}_{1c}a^{\dagger}_{2d}+ \beta_2\cdot a^{\dagger}_{1}  + \lambda_2\cdot a^{\dagger}_{2}}|0 \rangle\,.
\label{NES7X}
\ee
Inserting again the completeness relation in \eqref{completeness} with $n=2$ we get
\begin{align}
\mathcal{I}_{\zeta_1,\zeta_2}(\beta_i,\lambda_i)=\int\frac{d^Dv_1}{\pi^D}\int\frac{d^Dv_2 }{\pi^D}\mathrm{exp}\{-V_b\Delta^{bc}\bar{V}_{c}+V_b\bar\Gamma^b+\Gamma_b\bar V^b\},
\end{align}
where we defined
\begin{align}
V^b=(\bar v_1^b,v_2^b),\quad\bar{V}^b=\binom{v_1^b}{\bar v_2^b},\quad\Gamma^b=( \beta_1^b,\lambda_2^b),\quad \bar{\Gamma}^M=\binom{ \beta_2^b}{\lambda_1^b},
\end{align}
and
\begin{align}
\Delta^{bc}=\left(\begin{matrix}&\eta^{bc}&\zeta^{bc}_2\\&-\zeta^{bc}_1&\eta^{bc}\end{matrix}\right).
\end{align}
Therefore we can compute the Gaussian integral
\begin{align}
\nonumber \mathcal{I}_{\zeta_1,\zeta_2}(\beta_i,\lambda_i)&=\int\frac{d^Dv_1}{\pi^D}\int\frac{d^Dv_2 }{\pi^D}\mathrm{exp}\{-(V-\Gamma\Delta^{-1})\Delta(\bar{V}-\Delta^{-1}\bar{\Gamma})+\Gamma\Delta^{-1}\bar\Gamma\}\\&=({\mathrm{det}}\,\Delta)^{-1}\mathrm{exp}\{\Gamma\Delta^{-1}\bar\Gamma\}.
\end{align}
It is convenient to define
\begin{align}
M^{bc}:=(\zeta_2\cdot\zeta_1)^{bc},\qquad N^{bc}:=(\zeta_1\cdot\zeta_2)^{bc}=M^{cb},
\end{align}
in terms of which 
\begin{align}
(\mathrm{det}\,\Delta)^{-1}=\prod_{n=1}^{\infty}\mathrm{exp}\bigg(\frac{(-)^n}{n}(M^n)^b{}_b\bigg),
\end{align}
and $\Delta^{-1}_{bc}=\sum_{n=0}^\infty  (X^n)_{bc}$ 
with 
\begin{align}
   &(X^{2n})^{bc}=(-1)^n\left(\begin{matrix}&(M^n)^{bc} & 0\\ & 0 & (N^n)^{bc}\end{matrix}\right),\\&(X^{2n+1})^{bc}=(-1)^n\left(
   \begin{matrix}
       0 &(M^n)^{b}{}_{d}\zeta_2^{dc} \\-(N^n)^{b}{}_{d}\zeta_1^{dc} & 0\end{matrix}\right).
\end{align}
Note that by definition $(M^0)^{bc}=(N^0)^{bc}:=\eta^{bc}$.
The expression of the correlation function \eqref{NES7X} is given by 
\begin{align}
& {\cal{I}}_{\zeta_1,\zeta_2}(\beta_i,\lambda_i)=   \prod_{m=1}^\infty\exp \bigg(   \frac{(-)^m}{m}   (M^m)^{b}_{\,\,b}\bigg) \exp \bigg[ \sum_{n=0}^\infty (-)^n \Big(\beta_1\cdot M^n\cdot \beta_2+\lambda_2\cdot N^n\cdot\lambda_1  \nonumber \\
 &+  \beta_1\cdot M^n\cdot\zeta_2\cdot \lambda_1 -\lambda_2\cdot  N^{n}\cdot\zeta_1\cdot\beta_2\Big) \bigg].
    \label{NSFINAL}
\end{align}
The generating function \eqref{3pt2HaZetgen}, that we rewrite here for convenience as follows
\be
{\cal A}_{\mathrm{2h}}^{(\mathrm{gen})}:=\bra{p_{1}}e^{c_1\zeta^1_{b_1 b_2} a^{b_1}_1 a_2^{b_2}}\oint {dz\over 4\alpha'\pi i\,z}\, {\partial\over \partial\gamma} \,e^{\gamma{\epsilon}{\cdot}i\partial \hat{X}+ik{\cdot}\hat{X}}(z)\big|_{\gamma\mapsto 0}\,e^{c_2\zeta^2_{b_1 b_2} a^{\dagger b_1}_1 a_2^{\dagger b_2}}\ket{p_2}\,,
\ee
can be written in terms of $\mathcal{I}_{\zeta_1,\zeta_2}$ as
\begin{align}
\label{equation}
{\cal A}_{\mathrm{2h}}^{(\mathrm{gen})}=\frac{1}{4\pi\alpha'}\int_0^{2\pi}d\tau\,\frac{\partial}{\partial\gamma}e^{2\alpha'\gamma\epsilon\cdot p_2}\mathcal{I}_{c_1\zeta_1,c_2\zeta_2}(\beta_i(\tau),\lambda_i(\tau))\big|_{\gamma\mapsto0},
\end{align}
where $\beta_i(\tau)$ and $\lambda_i(\tau)$ are given in \eqref{ident1} and \eqref{relparameters}. Explicitly,
\begin{align}
&\nonumber\mathcal{I}_{c_1\zeta_1,c_2\zeta_2}(\beta_i(\tau),\lambda_i(\tau))=\mathrm{exp}\bigg\{\sum_{n=0}^{\infty}(-c_1 c_2)^n\alpha'\bigg[6\gamma^2\epsilon\cdot M^n\cdot\epsilon-3k\cdot M^n\cdot k\\\nonumber &-4\gamma f_{bc}(M^n)^{bc}+\gamma\sqrt{2}\epsilon\cdot (c_2e^{-3i\tau}M^n\cdot\zeta_2 -c_1e^{3i\tau}\zeta_1\cdot M^n)\cdot k\bigg]\bigg\}\\&\prod_{m=1}^{\infty}\mathrm{exp}\bigg(\frac{(-c_1c_2)^m}{m}(M^m)^b{}_b\bigg),
\end{align}
where $f_{bc}=k_{[b}\epsilon_{c]}$ is the field strength and where we used the antisymmetry of $M^n\cdot\zeta_2$ and $N^n\cdot\zeta_1$. 
Performing the integral over $\tau$ in \eqref{equation} yields the result in \eqref{genf2h}. 
\section{Expectation values of spin operators}
In this appendix, using the explicit results obtained in appendix \ref{appA}, we derive the expectation values of the spin operator and its square for states on the leading Regge trajectory and on the tilted trajectory. We define
\begin{align}
  \hat{S}^{ab}:=P^a{}_cP^b{}_d\hat{J}^{cd},\qquad 
\hat{J}^{ab}=-i\sum_{n=1}^{\infty}a^{\dagger[a}_na_n^{b]}\,,\qquad P^a{}_b:=\delta^a_b+\frac{p^ap_b}{m^2},
\end{align}
and \be
\label{expvaluesLR}
\langle\hat{S}^{ab}\rangle_{\,\mathrm{X}}^{(s)}:=\langle \Psi_{\,\mathrm{X}}^{(s)}(p)|\hat{S}^{ab}|\Psi_{\,\mathrm{X}}^{(s)}(p)\rangle\,,\quad \langle \hat{S}^{ac}\hat{S}_c{}^b\rangle_{\,\mathrm{X}}^{(s)}:=\langle \Psi_{\,\mathrm{X}}^{(s)}(p)|\hat{S}^{ac}\hat{S}_c{}^b|\Psi_{\,\mathrm{X}}^{(s)}(p)\rangle,
\ee
where X$=\{\mathrm{LR},2\mathrm{h}\}$.
\subsection{One-harmonic states}\label{LRexVals}
 Introducing the differential operator
\begin{align}
\label{eq:operatorO}
\hat{\mathcal{O}}^{ab}:=\frac{\partial^2}{\partial\beta_{2a}\partial\beta_{1b}},
\end{align}
then it is clear from \eqref{CFLR} that the first expectation value in \eqref{expvaluesLR} can be computed as
\begin{align}
\label{eq:spinoneharm}
    \langle\hat{S}^{ab}\rangle_{\,\mathrm{LR}}^{(s)}=-i\,P^a{}_cP^b{}_d\frac{1}{s!}\bigg(\frac{\partial}{\partial c_1}\bigg)^s\bigg(\frac{\partial}{\partial c_2}\bigg)^s\hat{\mathcal{O}}^{[cd]}\mathcal{C}_{c_1\bar{\varepsilon},c_2\varepsilon}(\beta_i)\big|_{c_i,\beta_i\mapsto0}=is\,\varepsilon^{[a}\bar{\varepsilon}^{b]}.
\end{align}
In order to get the expectation value of the second operator in \eqref{expvaluesLR} we first find it convenient to rewrite the operator $\hat{J}^{ab}\hat{J}^{cd}$ in such a way that all the operators $a^{\dagger}$ are on the right and $a$ on the left using the oscillator algebra:
\begin{align}
a^{\dagger a}_1a^b_1a^{\dagger c}_1 a^d_1=(\eta^{ab}\eta^{cd}-\eta^{ab}a^d_1a^{\dagger c}_1-\eta^{dc}a^b_1a ^{\dagger a}_1-\eta^{da}a^b_1a^{\dagger c}_1+a^b_1a^d_1a^{\dagger a}_1a^{\dagger c}_1).
\end{align}
By the definition of the correlator in \eqref{CFLR} we see that the correct expression for the expectation value of the above operator is
\begin{multline}
\langle a^{\dagger a}_1a^b_1a^{\dagger c}_1 a^d_1\rangle_{\mathrm{LR}}^{(s)}=\frac{1}{s!}\bigg(\frac{\partial}{\partial c_1}\bigg)^s\bigg(\frac{\partial}{\partial c_2}\bigg)^s\bigg(\eta^{ab}\eta^{cd}-\eta^{ab}\hat{\mathcal{O}}^{cd}-\eta^{dc}\hat{\mathcal{O}}^{ab}\\-\eta^{da}\hat{\mathcal{O}}^{cb}+\hat{\mathcal{O}}^{ab}\hat{\mathcal{O}}^{cd}\bigg)\mathcal{C}_{c_1\bar{\varepsilon},c_2\varepsilon}(\beta_i)\big|_{c_i,\beta_i\mapsto0}.
\end{multline}
Taking into account the antisymmetrization we see that
\begin{multline}
 \langle \hat{J}^{ab}\hat{J}^{cd}\rangle_{\mathrm{LR}}^{(s)}=-\frac{1}{s!}\bigg(\frac{\partial}{\partial c_1}\bigg)^s\bigg(\frac{\partial}{\partial c_2}\bigg)^s\bigg(\hat{\mathcal{O}}^{[ab]}\hat{\mathcal{O}}^{[cd]}-(\eta^{da}\hat{\mathcal{O}}^{cb}-\eta^{db}\hat{\mathcal{O}}^{ca}\\-\eta^{ca}\hat{\mathcal{O}}^{db}+\eta^{cb}\hat{\mathcal{O}}^{da})\bigg)\mathcal{C}_{c_1\bar{\varepsilon},c_2\varepsilon}(\beta_i)\big|_{c_i,\beta_i\mapsto0},
\end{multline}
yielding
\begin{align}
\langle \hat{J}^{ab}\hat{J}^{cd}\rangle_{\mathrm{LR}}^{(s)}=s[-(s+1)\varepsilon^{[a}\bar\varepsilon^{b]}\varepsilon^{[c}\bar\varepsilon^{d]}+\bar\varepsilon^b\varepsilon^{[d}\eta^{c]a}-\bar\varepsilon^a\varepsilon^{[d}\eta^{c]b}].
\end{align}
Correspondingly,
\begin{align}
\label{eq:LRspinexp}
\langle \hat{S}^{ab}\hat{S}^{cd}\rangle_{\mathrm{LR}}^{(s)}=s[-(s+1)\varepsilon^{[a}\bar\varepsilon^{b]}\varepsilon^{[c}\bar\varepsilon^{d]}+\bar\varepsilon^b\varepsilon^{[d}P^{c]a}-\bar\varepsilon^a\varepsilon^{[d}P^{c]b}],
\end{align}
and 
\begin{align}
\label{eq:LRspin2exp}
\langle \hat{S}^{ac}\hat{S}_{c}{}^{b}\rangle_{\mathrm{LR}}^{(s)}=-s[\bar\varepsilon^a\varepsilon^b(D+s-4)+(s-1)\varepsilon^a\bar\varepsilon^b+P^{ab}],
\end{align}
and
\begin{align}
\langle \hat{S}^{ac}\hat{S}_{ca}\rangle_{\mathrm{LR}}^{(s)}=-2s(s+D-3)\stackrel{D=4}{=}-2s(s+1).
\end{align}

\subsection{Tilted trajectory}
\label{TwHexVals}
We start by explicitly writing the correlation function in \eqref{NSFINAL} for the  factorised state in \eqref{eq:tworowfactorepsilon}. Using equation \eqref{eq:factpower}, that we rewrite here for $\zeta_2=\zeta $ and $\zeta_1=\bar{\zeta}$,
\begin{align}
(M^n)^{ab}=(-n_\zeta)^{n-1}(\zeta\cdot\bar{\zeta})^{ab},\qquad n>0,
\end{align}
with $n_\zeta=-\frac{1}{2}\mathrm{Tr}(\zeta\cdot\bar{\zeta})$ we get 
\begin{align}
\label{eq:correlatorresummed}
\nonumber&\mathcal{I}_{c_1\bar{\zeta},c_2\zeta}(\beta_i,\lambda_i)=g(c_1,c_2,n_\zeta)\,\mathrm{exp}\bigg\{\beta_1\cdot\beta_2+\lambda_1\cdot\lambda_2+c_2\beta_1\cdot\zeta\cdot\lambda_1-c_1\lambda_2\cdot\bar{\zeta}\cdot\beta_2\\&-f(c_1,c_2,n_\zeta)(\beta_1\cdot\zeta\cdot\bar\zeta\cdot\beta_2+\lambda_1\cdot\zeta\cdot\bar\zeta\cdot\lambda_2+c_2\beta_1\cdot\zeta\cdot\bar\zeta\cdot\zeta\cdot\lambda_1-c_1\lambda_2\cdot\bar\zeta\cdot\zeta\cdot\bar\zeta\cdot\beta_2)\bigg\},
\end{align}
with 
\begin{align}
g(c_1,c_2,n_\zeta):=\frac{1}{(1-c_1c_2n_\zeta)^2},\qquad f(c_1,c_2,n_\zeta):=\frac{c_1c_2}{1-c_1c_2n_\zeta}.
\end{align}
Together with the operator in \eqref{eq:operatorO}, we introduce also 
\begin{align}
\hat{\mathcal{T}}^{ab}:=\frac{\partial^2}{\partial\lambda_{2a}\partial\lambda_{1b}},
\end{align}
in terms of which 
\begin{align}
\langle\hat{S}^{ab}\rangle_{\mathrm{2h}}^{(s)}=\langle\hat{S}^{ab}_1\rangle_{\mathrm{2h}}^{(s)}+\langle\hat{S}^{ab}_2\rangle_{\mathrm{2h}}^{(s)},
\end{align}
where the single oscillator angular momentum are $\hat{J}_i^{ab}=a_i^{\dagger[a}a_i^{b]}$ and where
\begin{align}
&\langle\hat{S}^{ab}_1\rangle_{\mathrm{2h}}^{(s)}=-i|N(\zeta,s)|^2P^a{}_c P^{b}{}_d\bigg(\frac{\partial}{\partial c_1}\bigg)^s\bigg(\frac{\partial}{\partial c_2}\bigg)^s\hat{\mathcal{O}}^{[cd]}\mathcal{I}_{c_1\bar{\zeta},c_2\zeta}(\beta_i,0)\big|_{c_i,\beta_i\mapsto0},\\&\langle\hat{S}^{ab}_2\rangle_{\mathrm{2h}}^{(s)}=-i|N(\zeta,s)|^2P^a{}_c P^{b}{}_d\bigg(\frac{\partial}{\partial c_1}\bigg)^s\bigg(\frac{\partial}{\partial c_2}\bigg)^s\hat{\mathcal{T}}^{[cd]}\mathcal{I}_{c_1\bar{\zeta},c_2\zeta}(0,\lambda_i)\big|_{c_i,\lambda_i\mapsto0},
\end{align}
are the contributions of the two harmonics. Because of the structures in \eqref{eq:correlatorresummed} they turn out to be equal and we get
\begin{align}
\label{eq:spintwoharm}
\langle\hat{S}^{ab}\rangle_{\mathrm{2h}}^{(s)}=-\frac{i}{n_\zeta}s(\zeta\cdot\bar{\zeta})^{[ab]}.
\end{align}
For the operator spin squared operator we proceed as for the leading Regge case and we find 
\begin{align}
\langle \hat{J}^{ab}\hat{J}^{cd}\rangle_{\mathrm{2h}}^{(s)}=\langle \hat{J}^{ab}_1\hat{J}^{cd}_1\rangle_{\mathrm{2h}}^{(s)}+\langle \hat{J}^{ab}_2\hat{J}^{cd}_2\rangle_{\mathrm{2h}}^{(s)}+\langle \hat{J}^{ab}_1\hat{J}^{cd}_2\rangle_{\mathrm{2h}}^{(s)}+\langle \hat{J}^{ab}_2\hat{J}^{cd}_1\rangle_{\mathrm{2h}}^{(s)},
\end{align}
with 
\begin{multline}
 \langle \hat{J}^{ab}_1\hat{J}^{cd}_1\rangle_{\mathrm{2h}}^{(s)}=-|N(\zeta,s)|^2\bigg(\frac{\partial}{\partial c_1}\bigg)^s\bigg(\frac{\partial}{\partial c_2}\bigg)^s\bigg(\hat{\mathcal{O}}^{[ab]}\hat{\mathcal{O}}^{[cd]}-(\eta^{da}\hat{\mathcal{O}}^{cb}-\eta^{db}\hat{\mathcal{O}}^{ca}\\-\eta^{ca}\hat{\mathcal{O}}^{db}+\eta^{cb}\hat{\mathcal{O}}^{da})\bigg)\mathcal{I}_{c_1\bar{\zeta},c_2\zeta}(\beta_i,0)\big|_{c_i,\beta_i\mapsto 0},
\end{multline}
\begin{multline}
 \langle \hat{J}^{ab}_2\hat{J}^{cd}_2\rangle_{\mathrm{2h}}^{(s)}=-|N(\zeta,s)|^2\bigg(\frac{\partial}{\partial c_1}\bigg)^s\bigg(\frac{\partial}{\partial c_2}\bigg)^s\bigg(\hat{\mathcal{T}}^{[ab]}\hat{\mathcal{T}}^{[cd]}-(\eta^{da}\hat{\mathcal{T}}^{cb}-\eta^{db}\hat{\mathcal{T}}^{ca}\\-\eta^{ca}\hat{\mathcal{T}}^{db}+\eta^{cb}\hat{\mathcal{T}}^{da})\bigg)\mathcal{I}_{c_1\bar{\zeta},c_2\zeta}(0,\lambda_i)\big|_{c_i,\lambda_i\mapsto 0},
\end{multline}
and
\begin{align}
 \langle \hat{J}^{ab}_1\hat{J}^{cd}_2\rangle_{\mathrm{2h}}^{(s)}=-|N(\zeta,s)|^2\bigg(\frac{\partial}{\partial c_1}\bigg)^s\bigg(\frac{\partial}{\partial c_2}\bigg)^s\bigg(\hat{\mathcal{O}}^{[ab]}\hat{\mathcal{T}}^{[cd]}\bigg)\mathcal{I}_{c_1\bar{\zeta},c_2\zeta}(0,\lambda_i)\big|_{c_i,\lambda_i\mapsto 0}.
\end{align}
Putting everything together and considering the spin squared tensor we get
\begin{align}
&\langle \hat{S}^{ac}_1\hat{S}_{1c}{}^{b}\rangle_{\mathrm{2h}}^{(s)}+(1\leftrightarrow2)=s\bigg[\frac{1}{n_\zeta}(\zeta\cdot\bar{\zeta})^{ba}(D+s-4)+\frac{1}{n_\zeta}(\zeta\cdot\bar{\zeta})^{ab}(s-1)-2P^{ab}\bigg],\\&
\langle \hat{S}^{ac}_1\hat{S}_{2c}{}^{b}\rangle_{\mathrm{2h}}^{(s)}+(1\leftrightarrow2)=-\frac{s}{n_\zeta}(\zeta\cdot\bar{\zeta})^{(ab)},
\end{align}
yielding
\begin{align}
\langle \hat{S}^{ac}\hat{S}_{c}{}^{b}\rangle_{\mathrm{2h}}^{(s)}=\bigg[\frac{1}{n_\zeta}(\zeta\cdot\bar{\zeta})^{ba}(D+s-5)+\frac{1}{n_\zeta}(\zeta\cdot\bar{\zeta})^{ab}(s-2)-2P^{ab}\bigg],
\end{align}
and thus
\begin{align}
\langle \hat{S}^{ab}\hat{S}_{b}{}^{a}\rangle_{\mathrm{2h}}^{(s)}=-4s(s+D-4)\stackrel{D=4}{=}-4s^2.
\end{align}

\section{Three-point amplitudes with arbitrarily excited string states in the DDF formalism}

\subsection{Generating amplitude}
In the bosonic string at  critical dimension $D=26$ a complete set of physical states is provided by the DDF states~\cite{DelGiudice:1971yjh}
\beq
\label{notab}
 {\cal A}^{\mu}_{-n} := \oint \frac{dz}{2\pi i} i\partial X^{\mu}(z)\, e^{-in\sqrt{2\alpha'}q{\cdot}X}
~.
\eeq
which satisfy the oscillator algebra  
\be
[{\cal A}_{n}^{\mu},{\cal A}_{m}^{\nu}]=n\delta^{\mu\nu}\delta_{n+m,0}\,.
\ee
The generic vertex operator, corresponding to a physical state, can be realised by computing the action\footnote{here the action is computed in the sense of the operator-product-expansion.} of DDF creation operators on the tachyon vertex operator
\beq
\label{notad}
{\cal V}_{\{n,g_{n}\}}(H,p;z)=H_{(\mu_{1})_{g_{1}}(\mu_{2})_{g_{2}}...(\mu_{n})_{g_{n}}}\prod_{n}{1\over \sqrt{n^{g_{n}} g_{n}!}} \prod_{r=1}^{g_n} {\cal A}_{-n}^{\mu_{n}^{r}}\, :e^{i\sqrt{2\alpha'}p_{t}{\cdot}X}(z): 
\eeq
with the following constraints
\be
q{\cdot}{\cal A}_{-n}=0\,,\quad q^{2}=0\,,\quad \alpha'p_{t}^{2}=1\,,\quad \alpha'M^{2}=\sum_{n=1}^{N}ng_{n}-1\,,\quad p^{\mu}=p^{\mu}_{t}-q^{\mu}\sum_{n=1}^{N}ng_{n}
\label{constDDF}
\ee 
and where the generic polarization tensor contains {$g_{1}$ symmetric  indices, $\mu_{1}$, $g_2$ symmetric indices $\mu_2$, etc}. 
\be\label{genpoltens}
H_{(\mu_{1})_{g_{1}}(\mu_{2})_{g_{2}}\ldots} :=H_{(\mu^{1}_{1}...\mu^{g_{1}}_{1})(\mu^{1}_{2}...\mu^{g_{2}}_{2})\ldots}\,,\quad (\mu_{r})_{g_{r}}:=(\mu_{r}^{1}\mu_{r}^{2}....\mu_{r}^{g_{r}}) ~,
\ee
Any given state
\be
|\{n,g_{n}\},H,p\rangle=\lim_{z\rightarrow 0}{\cal V}_{\{n,g_{n}\}}(H,p;z)
\ee
can be normalised by requiring 
\be
\,^{*}\big\langle \{n,g_{n}\},H,p\big|\{n,g_{n}\},H,p\big\rangle=1\,\,\Rightarrow \,\,Tr(H^{T}H^{*})=1
\ee
In~\cite{Ademollo:1974kz} the three-point amplitude involving three arbitrary DDF vertex operators was constructed. Here we follow a bit different path by first constructing the generating function involving three arbitrary DDF states. In fact, using coherent vertex operators \cite{Skliros:2011si,Hindmarsh:2010if} of the form
\beq
\label{cohaa}
{\cal V}_{coh}(\{\epsilon_{n}\};z)=\,:e^{\sum_{n}\epsilon_{n}{\cdot}{\cal A}_{-n}}:\, :e^{i\sqrt{2\alpha'}p_{t}{\cdot}X}(z): 
\eeq
$$
=\exp{\left(\sum_{n,m}\frac{\widehat{\epsilon}_{n}{\cdot}\widehat{\epsilon}_{m}}{2}{\cal S}_{n,m}e^{-i(n{+}m)\sqrt{2\alpha'}q{\cdot}X}{+}\sum_{n}\widehat{\epsilon}_{n}{\cdot}{\cal P}_{n}e^{-in\sqrt{2\alpha'}q{\cdot}X}{+}i \sqrt{2\alpha'}p_{t}{\cdot}X\right)}(z)
$$
which are made of the following polynomial structures  
\beq
\label{cohac}
{\cal P}^{\mu}_{n}(z) := \sum_{k=1}^{n} \frac{i\partial^{k} X^{\mu}}{(k{-}1)!}{\cal Z}_{n{-}k}(a_{s}^{(n)})~, ~~
a_s^{(n)} :=-in \frac{q{\cdot}\partial^{s}X}{(s{-}1)!}
~,
\eeq
\beq
\label{cohad}
{\cal S}_{n,m} := \sum_{r=1}^{m}r\,{\cal Z}_{n{+}r}(a_{s}^{(n)}){\cal Z}_{m{-}r}(a_{s}^{(m)})\,,\quad  {\cal Z}_{n}(x)=\frac{1}{2\pi i}\oint \frac{dw}{w^{n{+}1}} e^{\sum_{s=1}^{\infty}\frac{x}{s}w^{s}}\,,
\eeq
with polarisation sources 
\be
\widehat{\epsilon}^{\mu}=\epsilon^{\mu}-2\alpha'\epsilon{\cdot}p\,q^{\mu}\,,
\label{RotPol}
\ee
one can obtain the generating function of three point amplitudes by computing the amplitude of three coherent states 
\be
A_{gen}^{DDF}:=\Big\langle c(z_{1}){\cal V}_{coh}(\{\epsilon^{(1)}_{n}\};z_{1}) c(z_{2}){\cal V}_{coh}(\{\epsilon^{(2)}_{n}\};z_{2}) c(z_{3}){\cal V}_{coh}(\{\epsilon^{(3)}_{n}\};z_{3}) \Big\rangle
\ee
where the fields $c(z_{j})$ are the usual ghost operators. Using the constraints \eqref{constDDF}, with an additional collinear condition for the DDF reference momenta $q^{\mu}_{1}\propto q^{\mu}_{2} \propto q^{\mu}_{3}$, and the contraction rule
\be
\langle X^{\mu}(z_{1}) X^{\nu}(z_{2}) \rangle=- \eta^{\mu \nu}\log(z_{1}-z_{2})
\ee
the resulting expression of the generating function of three point amplitudes is given by
\begin{equation}
A_{gen}^{DDF}=\exp \sum_{r=1}^3\Bigg[ \sum_{n=1}^\infty  \sqrt{2\alpha'}\,{\widehat{\epsilon}}_n^{(r)}{\cdot}p_{r+1} N_n^{(r)}  +\sum_{n,m=1}^\infty  \Bigg(   {{\widehat{\epsilon}}_n^{(r)}{\cdot}{\widehat{\epsilon}}_m^{(r)}\over 2}N_{nm}^{(r,r)} + {\widehat{\epsilon}}^{(r)}_n{\cdot}{\widehat{\epsilon}}^{(r+1)}_mN_{nm}^{(r,r+1)}\Bigg)  \Bigg]
\label{FR1X}
\end{equation}
where~
\footnote{since the generating function is manifestly symmetric under cyclic transformations, and there are three states, we use a notation with implicit cyclic symmetry for scalar products of the form  $q_{r}{\cdot}p_{r+2}=q_{r}{\cdot}p_{r-1}$. }
\begin{equation}
N_n^{(r)}= {(-)^{n+1}\over (n-1)!}( 1+ n2\alpha'q_r{\cdot}p_{r+1})_{n-1}  
\label{FR2X}
\end{equation}
\begin{equation}
N_{nm}^{(r,s)} = - \frac{nm}{m+ n 2\alpha' q_{r}{\cdot}p_{s}}2\alpha' q_r{\cdot}p_{s+1}2\alpha' q_r{\cdot}p_{s-1} N_n^{(r)} N_m^{(s)}\,.
\label{FR3XXX}
\end{equation}
and  $a_n= \frac{\Gamma (a+n)}{\Gamma (a)}$ is the Pochhammer symbol.

Given the coherent vertex operator \eqref{cohaa}, a generic vertex operator \eqref{notad} is obtained by the following derivative projection 
\be
{\cal V}_{\{n,g_{n}\}}(H,p;z)=H_{(\mu_{1})_{g_{1}}(\mu_{2})_{g_{2}}...(\mu_{n})_{g_{n}}}\prod_{n}{1\over \sqrt{n^{g_{n}} g_{n}!}} \prod_{r=1}^{g_n} {\partial \over \partial \epsilon_{n, \mu_{n}^{g_{r}}}}{\cal V}_{coh}(\{\epsilon_{n}\};z)\big|_{\epsilon_{n}=0}
\label{DprojCohe}
\ee 
and with the same logic, from the generating function of amplitudes \eqref{FR1X}, one can extract the amplitude of three generic states taking a similar derivative projection for all the states
\be
A(H^{(1)},H^{(2)},H^{(3)})=\prod_{j=1}^{3}H^{(j)}_{(\mu_{1})_{g_{1}}...(\mu_{n_{j}})_{g_{n_{j}}}}\prod_{n_{j}}{1\over \sqrt{n_{j}^{g_{n_{j}}} g_{n_{j}}!}} \prod_{r_{j}=1}^{g_{n_{j}}} {\partial \over \partial \epsilon^{\hspace{0cm}^{(j)}}_{\vspace{-1mm}n_{j}, \mu_{n_{j}}^{g_{r_{j}}}}}A_{gen}^{DDF}\Bigg|_{\epsilon^{(1,2,3)}_{n}=0}
\label{DprojGenA}
\ee
For the case in which the state in position $2$ is massless, meaning that there is only the harmonic $n=1$ with excitation number $g_{1}=1$, the polarisation tensor is restricted to 
\be
H^{(2)}_{{(\mu_{1})_{g_{1}}(\mu_{2})_{g_{2}}...(\mu_{n_{2}})_{g_{n_{2}}}}}\Rightarrow H^{(2)}_{\mu}
\ee
and according to \eqref{RotPol}, we can introduce the final polarization vector as 
\be
{\cal E}_\mu=H_\mu^{(2)}-2\alpha'k{\cdot}H^{(2)}q_{2,\mu}
\ee
which, as described by the set of constrains \eqref{constDDF}, represents the polarization vector of a massless state transverse to its momentum $k^{\mu}=p_{t,2}^\mu-q_2^\mu$. The latter has the same degrees of freedom of the polarization vector $\ep^\mu$ used in the previous sections.

The generating function in the present case is given by
\be
 A^{DDF}_{gen,{\cal E}}=\left( \sqrt{2\alpha'} {\cal E}{\cdot}p_3 +{\cal E}{\cdot}{\widehat{\epsilon}}_m^{(3)} N_{1m}^{23} +  
{\widehat{\epsilon}}_n^{(1)}{\cdot}{\cal E} N_{n1}^{12} \right)
\label{C1X}
\ee
$$
\exp \Bigg[ \sum_{n=1}^\infty \sum_{r=1,3} \sqrt{2\alpha'}\,{\widehat{\epsilon}}_n^{(r)}{\cdot}p_{r+1} N_n^{(r)}  +\sum_{n,m=1}^\infty \sum_{r=1}^3 \Bigg(  \frac{1}{2} {\widehat{\epsilon}}_n^{(r)}{\cdot}{\widehat{\epsilon}}_m^{(r)}N_{nm}^{(r,r)} + {\widehat{\epsilon}}^{(r)}_n{\cdot}{\widehat{\epsilon}}^{(r+1)}_mN_{nm}^{(r,r+1)}\Bigg)  \Bigg]
$$
In order to compute $N^{(r)}_n$ and $N^{(r,s)}_{nm}$ appearing in \eqref{FR2X} and \eqref{FR3XXX} we need to fix the kinematic of three-point amplitude with a massless state and two equal states. For consistency, in this case we need to have complex momenta. We choose the momenta of the three external states  to have the following  coordinates:
\begin{align}
&p_1^\mu = (-E,- i \frac{k}{2}, p, O_{22}, - \frac{k}{2})~; p_2^\mu = k^\mu = (0, ik, O_{23}, k)~;p_3^\mu = (E, - i \frac{k}{2}, -p, O_{22}, - \frac{k}{2})
\nonumber \\
& k^2=0 ~~;~~\alpha' M^2 \equiv -\alpha' p_1^2=-\alpha'p_3^2= \alpha' (E^2-p^2)= -1+
\sum_i n_i
\label{ARB1}
\end{align}
They satisfy momentum conservation:
\begin{equation}
p_1+p_3+k=0
    \label{momcons}
\end{equation}
We also need  to fix the coordinates of the momenta of the three tachyons and the three auxiliary massless momenta. They are given by
\begin{align}
& p_{t,1}= (-E,-i (\frac{k}{2}- 
\frac{n}{2\alpha'k}),p, O_{22}, -\frac{k}{2} - \frac{n}{2\alpha'k})~~;~~\alpha' p_{t,1}^2 =1 \nonumber \\
& p_{t,2}= (0, \frac{i}{2}( k - \frac{1}{4\alpha' k}), O_{23}, k+ \frac{1}{4\alpha' k}) ~~;~~ \alpha'p_{t,2}^2=1 \nonumber \\
& p_{t,3}= (E, -i ( \frac{k}{2} - \frac{n}{2\alpha' k}), -p, O_{22}, - \frac{k}{2} - \frac{n}{2\alpha' k}) ~~;~~\alpha' p_{t,3}^2=1 \nonumber \\
&q_1 =q_3 =-2 q_2 =-\frac{1}{2\alpha' k} (0, -i, 0_{23}, 1)~~;~~q_1^2=q_2^2=q_3^2 =0 
    \label{p1p2p3}
\end{align}
where $n=\sum_i n_i$.
Using the previous expressions one gets
\begin{align}
&2\alpha' p_1 q_2 = - \frac{1}{2}~~;~~2\alpha' p_3 q_2 = - \frac{1}{2} ~~;~~2\alpha' p_1 q_3= 1~~;~~2\alpha' p_2 q_3 =-2 \\
& 2\alpha' p_2 q_1 =-2 ~~;~~ 2\alpha' p_3 q_1=1 \Longrightarrow  2\alpha' q_{i} (p_1+p_2+p_3)=0\,\,\,i=1,2,3
\label{ARB9X}
\end{align}
They can be used in \eqref{C1X} to arrive at the final formula for the three-point coupling involving two identical arbitrarily excited states and a massless state:
\be
A^{DDF}_{gen,{\cal E}}= \left[\sqrt{2\alpha'} {\cal E}{\cdot}p_3 + \sum_{m=1}^{\infty} {\Gamma(2m)\over \Gamma(m)^{2}}{1\over 2m{-}1}\Big((-)^{m+1}{\cal E}{\cdot}\widehat{\epsilon}_{m}^{(3)} +\widehat{\epsilon}_{m}^{(1)}{\cdot}{\cal E}  \Big) \right]
 \label{ARB16}
\ee
$$
\exp \Bigg( \sqrt{2\alpha'} \sum_{n=1}^\infty\frac{  \Gamma (2n)}{n\Gamma (n)^2}  \Big( {\widehat{\epsilon}}_n^{(1)}{\cdot}k+
(-)^{n}  {\widehat{\epsilon}}_n^{(3)}{\cdot}k \Big) + 2\sum_{n,m=1}^\infty \frac{(-)^{m+1} }{n+m}\frac{\Gamma (2n)}{\Gamma (n)^2} \frac{\Gamma (2m)}{\Gamma (m)^2} 
  {\widehat{\epsilon}}_n^{(1)}{\cdot}{\widehat{\epsilon}}_m^{(3)}
$$
$$
 +  \sum_{n,m=1}^\infty \frac{1}{m+n} \frac{\Gamma (2n)}{\Gamma (n)^2 } \frac{\Gamma (2m)}{\Gamma (m)^2 }   \Big(   {\widehat{\epsilon}}_n^{(1)}{\cdot}{\widehat{\epsilon}}_m^{(1)}  +  (-)^{n+m} \,{\widehat{\epsilon}}_n^{(3)}{\cdot}{\widehat{\epsilon}}_m^{(3)}\Big) \Bigg)
$$

\subsection{Three point amplitudes with excited states $({\cal A}_{-1})^N$}

The first case we study is the 
three-point coupling of two equal states with only the first harmonic excited and a massless state. The generating function  
\eqref{ARB16} can be restricted to 
\be
A^{DDF}_{gen,{\cal E}}=\left( \sqrt{2\alpha'} {\cal E}{\cdot}p_3 + {\cal E}{\cdot}\widehat{\epsilon}_{1}^{(3)} +\widehat{\epsilon}_{1}^{(1)}{\cdot}{\cal E}\right)  
\label{GenFunLReggwTr}
\ee
$$
\exp\Big( \sqrt{2\alpha'}\,{\widehat{\epsilon}}_1^{(1)}{\cdot}k- \sqrt{2\alpha'}\,{\widehat{\epsilon}}_1^{(3)}{\cdot}k + {\widehat{\epsilon}}_1^{(1)}{\cdot}{\widehat{\epsilon}}_1^{(3)} +{1\over 2}{\widehat{\epsilon}}_1^{(1)}{\cdot}{\widehat{\epsilon}}_1^{(1)}+{1\over 2}{\widehat{\epsilon}}_1^{(3)}{\cdot}{\widehat{\epsilon}}_1^{(3)}  \Big)\,.
$$
Using equation \eqref{DprojGenA}, which specialised to the present case is
\be
{1\over \sqrt{g_{1}!}}H^{(1)}_{(\mu_{1})_{g_{1}}}\prod_{r_{1}=1}^{g_{1}}{\partial\over \partial \epsilon^{(1)}_{1,\mu_{1}^{r_{1}}}}   {1\over \sqrt{g_{1}!}}H^{(3)}_{(\nu_{1})_{g_{1}}}\prod_{r_{3}=1}^{g_{1}}{\partial\over \partial \epsilon^{(3)}_{1,\nu_{1}^{r_{3}}}}A^{DDF}_{gen,{\cal E},n=1}\Bigg|_{\epsilon^{(1)}_{n},\epsilon^{(3)}_{n}=0}\,,
\label{GenSpecLead}
\ee 
taking the polarizations $H^{(1)}$ and $H^{(3)}$ with the same symmetries of the manifestly covariant leading Regge states, $i.e.$ totally symmetric, transverse and traceless 
\be
H^{(j)}_{(\mu_{1}^{1}...\mu_{1}^{g_{1}})}\,,\quad H^{(j)}_{(\mu_{1}^{1}...\mu_{1}^{g_{1}})}p_{j}^{\mu_{1}}=0\,,\quad H^{(j)}_{(\mu_{1}^{1}...\mu_{1}^{g_{1}})}\eta^{\mu^{1}_{1}\mu^{2}_{1}}=0\,,
\label{PolCondLedRe}
\ee
and also taking a subset of general polarisations given by 
\be
H^{(j)}_{(\mu_{1}^{1}...\mu_{1}^{g_{1}})}\,\,\Rightarrow\,\, \bigotimes_{r=1}^{g_{j}}\xi^{(j)}_{\mu_{r}}\,,\quad \xi^{(j)}{\cdot}\xi^{(j)}=0
\ee 
the expression \eqref{GenSpecLead} reduces to 
\be
A(\{\xi^{(1)}\}_{N},{\cal E},\{\xi^{(3)}\}_{N})={1\over N!}\left(\xi^{(1)}{\cdot}{\partial\over \partial \epsilon^{(1)}_{1}}\xi^{(3)}{\cdot}{\partial\over \partial \epsilon^{(3)}_{1}}\right)^{N}   A^{DDF}_{gen,{\cal E},n=1}\Bigg|_{\epsilon^{(1)}_{n},\epsilon^{(3)}_{n}=0}
\label{ReducGenF}
\ee
where it was used the mass level relation $N=\sum_{n}n g_{n}$ for leading Regge states where $N=g_{1}$.

Now using the identity\footnote{Such identity is symmetric under $1\leftrightarrow 3$.}
\be\label{lag}
{d^{N_{3}}\over dJ^{N_{3}}_{3}}{d^{N_{1}}\over dJ^{N_{1}}_{1}}\,e^{J_{1}O_{1}+J_{3}O_{3} +J_{1}J_{3} O_{1,3}}\Bigg|_{J_{1},J_{3}{=}0}=N_{1}! (O_{1,3})^{N_{1}} (O_{3})^{N_{3}{-}N_{1}} \,L_{N_{1}}^{(N_{3}{-}N_{1})}\left(-{O_{1}O_{3}\over O_{1,3}} \right)
\ee
with $L^{(a)}_{N}(x)$ the Laguerre polynomials 
\be\label{LagPoly}
L^{(a)}_{N}(x)=\sum_{r=0}^{N}(-)^{r} \begin{pmatrix}N+a\\N-r\end{pmatrix}{x^{r}\over r!}
\ee
we can separate \eqref{ReducGenF} in two main contributions\footnote{Given the conditions \eqref{PolCondLedRe}, due to the absence of the traces, we can directly drop the corresponding terms in the generating function \eqref{GenFunLReggwTr}, meaning $\widehat{\epsilon}^{(1)}_{n}{\cdot}\widehat{\epsilon}^{(1)}_{m}=\widehat{\epsilon}^{(3)}_{n}{\cdot}\widehat{\epsilon}^{(3)}_{m}=0$. }, the first one is given by
\be
{1\over N!}\left(\xi^{(1)}{\cdot}{\partial\over \partial \epsilon^{(1)}_{1}}\xi^{(3)}{\cdot}{\partial\over \partial \epsilon^{(3)}_{1}}\right)^{N} \sqrt{2\alpha'} {\cal E}{\cdot}p_3\, e^{ \sqrt{2\alpha'}\,{\widehat{\epsilon}}_1^{(1)}{\cdot}k- \sqrt{2\alpha'}\,{\widehat{\epsilon}}_1^{(3)}{\cdot}k + {\widehat{\epsilon}}_1^{(1)}{\cdot}{\widehat{\epsilon}}_1^{(3)} }\big|_{\epsilon^{(1)},\epsilon^{(3)}=0}
\label{Contr1LR}
\ee
$$
=\sqrt{2\alpha'}{\cal E}{\cdot}p_{3}(\xi^{(1)}{\cdot}\xi^{(3)})^{N}\,L_{N}^{(0)}\left(2\alpha' {\xi^{(1)}{\cdot}k\,\xi^{(3)}{\cdot}k\over \xi^{(1)}{\cdot}\xi^{(3)}} \right)\,,
$$
while the second contribution is 
\be
{1\over N!}\left(\xi^{(1)}{\cdot}{\partial\over \partial \epsilon^{(1)}_{1}}\xi^{(3)}{\cdot}{\partial\over \partial \epsilon^{(3)}_{1}}\right)^{N}({\cal E}{\cdot}\widehat{\epsilon}_{1}^{(3)} {+}\widehat{\epsilon}_{1}^{(1)}{\cdot}{\cal E}) \,e^{ \sqrt{2\alpha'}\,{\widehat{\epsilon}}_1^{(1)}{\cdot}k- \sqrt{2\alpha'}\,{\widehat{\epsilon}}_1^{(3)}{\cdot}k + {\widehat{\epsilon}}_1^{(1)}{\cdot}{\widehat{\epsilon}}_1^{(3)} }\big|_{\epsilon^{(1)},\epsilon^{(3)}=0}
\label{Contr2LR}
\ee
$$
={(N{-}1)!\over N!}\begin{pmatrix} N\\ 1 \end{pmatrix}\sqrt{2\alpha'}({\cal E}{\cdot }\xi^{(3)}\xi^{(1)}{\cdot}k-{\cal E}{\cdot }\xi^{(1)}\xi^{(3)}{\cdot}k)(\xi^{(1)}{\cdot}\xi^{(3)})^{N-1}L_{N-1}^{(1)}\left(2\alpha' {\xi^{(1)}{\cdot}k\,\xi^{(3)}{\cdot}k\over \xi^{(1)}{\cdot}\xi^{(3)}} \right)
$$
$$
=\sqrt{2\alpha'}\xi^{(3)}{\cdot}f{\cdot}\xi^{(1)}(\xi^{(1)}{\cdot}\xi^{(3)})^{N-1}L_{N-1}^{(1)}\left(2\alpha' {\xi^{(1)}{\cdot}k\,\xi^{(3)}{\cdot}k\over \xi^{(1)}{\cdot}\xi^{(3)}} \right)
$$
where in the last step we have introduced $f^{[\mu\nu]}:=k^{[\mu}{\cal E}^{\nu]}$.
Combining \eqref{Contr1LR} and \eqref{Contr2LR} together, one arrives at the final result of the amplitude \eqref{ReducGenF}:
\be
A(\{\xi^{(1)}\}_{N},{\cal E},\{\xi^{(3)}\}_{N})=\sqrt{2\alpha'}{\cal E}{\cdot}p_{3}(\xi^{(1)}{\cdot}\xi^{(3)})^{N}\,L_{N}^{(0)}\left(2\alpha' {\xi^{(1)}{\cdot}k\,\xi^{(3)}{\cdot}k\over \xi^{(1)}{\cdot}\xi^{(3)}} \right)+
\ee
$$
+\sqrt{2\alpha'}\xi^{(1)}{\cdot}f{\cdot}\xi^{(3)}(\xi^{(1)}{\cdot}\xi^{(3)})^{N-1}L_{N-1}^{(1)}\left(2\alpha' {\xi^{(1)}{\cdot}k\,\xi^{(3)}{\cdot}k\over \xi^{(1)}{\cdot}\xi^{(3)}} \right)
$$
The resulting amplitude, computed here in the DDF formalism, has the same structure of the amplitude computed in the covariant formalism\footnote{Remeber that from momentum conservation ${\cal E}{\cdot}p_3=-{\cal E}{\cdot}p_1$.} \eqref{amplitudeLRs}, where we have to identify ${\cal E}^\mu$ with $\ep^\mu$, $\xi_{\mu}^{(1)}$ with $\ep_{1,\mu}$ and $\xi_{\mu}^{(3)}$ with $\ep_{2,\mu}$ .

\subsection{Three point amplitudes with excited states $({\cal A}_{-1}{\cal A}_{-2})^N$}
The next non trivial extension of the previous case, is the computation of the coupling of two equal states, with the first two harmonics excited, and a massless state. 

The generating function \eqref{ARB16}, restricted to the relevant terms for the present situation, is given by
\be
A^{DDF}_{gen,{\cal E}}=\Big(\sqrt{2\alpha'}{\cal E}{\cdot}p_{3}+{\cal E}{\cdot}\widehat{\epsilon}_{1}^{(1)}+{\cal E}{\cdot}\widehat{\epsilon}_{1}^{(3)}+2{\cal E}{\cdot}\widehat{\epsilon}_{2}^{(1)}-2{\cal E}{\cdot}\widehat{\epsilon}_{2}^{(3)}\Big)
\label{AgenDDF}
\ee
$$
\times\exp\Big(\sqrt{2\alpha'}\,\widehat{\epsilon}_{1}^{(1)}{\cdot}k- \sqrt{2\alpha'}\,\widehat{\epsilon}_{1}^{(3)}{\cdot}k+3\sqrt{2\alpha'}\,\widehat{\epsilon}_{2}^{(1)}{\cdot}k+3\sqrt{2\alpha'}\,\widehat{\epsilon}_{2}^{(3)}{\cdot}k \Big)
$$
$$
\times\exp\Big(\widehat{\epsilon}_{1}^{(1)}{\cdot}\widehat{\epsilon}_{1}^{(3)}-4\widehat{\epsilon}_{1}^{(1)}{\cdot}\widehat{\epsilon}_{2}^{(3)}+4\widehat{\epsilon}_{2}^{(1)}{\cdot}\widehat{\epsilon}_{1}^{(3)}-18 \widehat{\epsilon}_{2}^{(1)}{\cdot}\widehat{\epsilon}_{2}^{(3)} \Big)
$$
and the amplitude with generic excited states containing only two harmonics can be extracted from \eqref{DprojGenA} as 
\be
{1\over 2^{K}(K!)^{2}}{\cal H}^{(1)}_{(\mu_{1})_{K}(\mu_{2})_{K}}\prod_{j=1}^{K}{\partial \over \partial \epsilon^{(1)}_{1\,\mu_{1}^{j}}}{\partial \over \partial \epsilon^{(1)}_{2\,\mu_{2}^{j}}}{\cal H}^{(3)}_{(\nu_{1})_{K}(\nu_{2})_{K}}\prod_{\ell=1}^{K}{\partial \over \partial \epsilon^{(3)}_{1\,\nu_{1}^{\ell}}}{\partial \over \partial \epsilon^{(3)}_{2\,\nu_{2}^{\ell}}}A^{DDF}_{gen,{\cal E}}\big|_{\epsilon^{(1,3)}=0}\,.
\ee
where we select $g_{1}=g_{2}=K$ and  use \eqref{AgenDDF}.

Now choosing a subset of the general polarizations, with the factorised form 
\be
 {\cal H}_{(\mu_{1})_{K}(\mu_{2})_{K}}\,\,\Rightarrow \,\, \bigotimes_{j=1}^{K}\,H_{\mu_{1}\mu_{2}}
\ee
and asking for a tensor with the same symmetries of \eqref{eq:gentwomodestate} as described in \eqref{eq:twomodestatephyscond}, the generating function reduces to 
\be
A(\{H^{(1)}\}_{K},{\cal E},\{H^{(3)}\}_{K})={1\over 2^{K}(K!)^{2}}\left({\partial \over \partial \epsilon^{(1)}_{1}}{\cdot}H^{(1)}{\cdot}{\partial \over \partial \epsilon^{(1)}_{2}}\right)^{K}
\ee
$$
\times \left({\partial \over \partial \epsilon^{(3)}_{1}}{\cdot}H^{(3)}{\cdot}{\partial \over \partial \epsilon^{(3)}_{2}}\right)^{K}A^{DDF}_{gen,{\cal E}}\big|_{\epsilon^{(1)},\epsilon^{(3)}=0}\,.
$$
The first three amplitudes with low level states are the following :
\begin{itemize}
\item $K=1$
\end{itemize}
\be
 A(\{H^{(1)}\}_{1},{\cal E},\{H^{(3)}\}_{1}) =\sqrt{2\alpha'} {\cal E}{\cdot}p_{3} \,Tr(H^{(1)}H^{(3)})-{3\over 2}\sqrt{2\alpha'}{\cal E }{\cdot}p_{3}\,2\alpha'\,k{\cdot}H^{(1)}{\cdot}H^{(3)}{\cdot}k +
\ee
$$
+\,2 \,Tr\left(H^{(1)}{\cdot}f{\cdot}H^{(3)}\right)
$$

\begin{itemize}
\item $K=2$
\end{itemize}   
\be\label{Ku2tharm}
A(\{H^{(1)}\}_{2},{\cal E},\{H^{(3)}\}_{2})=\sqrt{2\alpha'}{\cal E}{\cdot}p_{3}\Big(    {1\over 2}Tr\left(H^{(1)}H^{(3)}\right)^{2}+
\ee
$$
-{3\over 2}Tr\left( H^{(1)}H^{(3)}\right)2\alpha'k{\cdot}H^{(1)}H^{(3)}{\cdot}k+
$$
$$
+{9\over 8} \left(2\alpha'k{\cdot}H^{(1)}H^{(3)}{\cdot}k\right)^{2}+{1\over 2}Tr\left((H^{(1)}H^{(3)})^{2}\right)-{3\over 2} 2\alpha'k{\cdot}(H^{(1)}H^{(3)})^{2}{\cdot}k \Big)+
$$
$$
+2Tr\left(H^{(1)}H^{(3)}\right)\sqrt{2\alpha'}\,Tr\left(H^{(1)}fH^{(3)}\right)+
$$
$$
-3\,\sqrt{2\alpha'}\,Tr\left(H^{(1)}fH^{(3)}\right)\,2\alpha'k{\cdot}H^{(1)}{\cdot}H^{(3)}{\cdot}k+
$$
$$
+2 \,\sqrt{2\alpha'}Tr\left(H^{(1)}fH^{(3)}H^{(1)}H^{(3)}\right)
$$

\begin{itemize}
\item $K=3$
\end{itemize}   
\be
A(\{H^{(1)}\}_{3},{\cal E},\{H^{(3)}\}_{3})=\sqrt{2\alpha'}{\cal E}{\cdot}p_{3}\Bigg(-{1\over 6}\, Tr\left(H^{(1)}H^{(3)}\right)^{3}+
\ee
$$
+{3\over 4}\, Tr\left(H^{(1)}H^{(3)}\right)^{2}2\alpha'\,k{\cdot}H^{(1)}{\cdot}H^{(3)}{\cdot}k +
$$
$$
-{9\over 8}\,Tr\left(H^{(1)}H^{(3)}\right)\left(2\alpha' \,k{\cdot}H^{(1)}{\cdot}H^{(3)}{\cdot}k \right)^{2}+{9\over 16}\left(2\alpha' k{\cdot}H^{(1)}{\cdot}H^{(3)}{\cdot}k \right)^{3}+
$$
$$
-{1\over 2}\,Tr\left(H^{(1)}H^{(3)}\right)\, Tr\left((H^{(1)}H^{(3)})^{2}\right)+{3\over 4}\,  Tr\left((H^{(1)}H^{(3)})^{2}\right) 2\alpha'\,k{\cdot}H^{(1)}{\cdot}H^{(3)}{\cdot}k +
$$
$$
+{3\over 2} \, Tr\left(H^{(1)}H^{(3)}\right) 2\alpha'k{\cdot}(H^{(1)}H^{(3)})^{2}{\cdot}k -{9\over 4}\,2\alpha'k{\cdot}H^{(1)}{\cdot}H^{(3)}{\cdot}k 2\alpha'k{\cdot}(H^{(1)}H^{(3)})^{2}{\cdot}k+
$$
$$
-{1\over 3}\, Tr\left((H^{(1)}H^{(3)})^{3}\right)+{3\over 2}\,2\alpha'k{\cdot}(F^{(1)}F^{(3)})^{3}{\cdot}k \Bigg)+
$$
$$
+\sqrt{2\alpha'}Tr\left(H^{(1)}f H^{(3)}\right){Tr\left(H^{(1)}H^{(3)}\right)^{2}}\times
$$
$$
\times \left(1- 3\, 2\alpha'{k{\cdot}H^{(1)}{\cdot}H^{(3)}{\cdot}k\over Tr\left(H^{(1)}H^{(3)}\right)}+\left(2\alpha'{3\over 2}{k{\cdot}H^{(1)}{\cdot}H^{(3)}{\cdot}k\over Tr\left(H^{(1)}H^{(3)}\right)}\right)^{2} \right) +
$$
$$
+2\sqrt{2\alpha'}Tr\left(H^{(1)}f H^{(3)}H^{(1)}H^{(3)}\right){Tr\left(H^{(1)}H^{(3)}\right)}\left(1-2\alpha'{3\over 2}{k{\cdot}H^{(1)}{\cdot}H^{(3)}{\cdot}k\over Tr\left(H^{(1)}H^{(3)}\right)} \right)+
$$
$$
+\sqrt{2\alpha'}Tr\left(H^{(1)}f H^{(3)}\right){Tr\left((H^{(1)}H^{(3)})^{2}\right)}\left(1-3\,2\alpha'{k{\cdot}(H^{(1)}{\cdot}H^{(3)})^{2}{\cdot}k\over Tr\left((H^{(1)}H^{(3)})^{2}\right)} \right) 
$$
$$
+ 2\,\sqrt{2\alpha'}Tr\left(H^{(1)}f H^{(3)}(H^{(1)}H^{(3)})^{2} \right)
$$

For generic values of $K$, one can find the following expression
 \begin{align}
 &A(\{H^{(1)}\}_{K},{\cal E},\{H^{(3)}\}_{K})=  \sqrt{2\alpha '} {\cal E}{\cdot}p_3  \prod_{\ell=1}^K \Bigg( \sum_{n_{\ell}=0}^\infty {(-)^{\ell n_{\ell}}\over n_{\ell}}  {\cal C}_{\ell}^{n_{\ell}}\Bigg)  \delta_{\sum_{i=1}^K \ell \,n_\ell\,;\,K} \nonumber \\
 & +2\sqrt{2\alpha'} \sum_{k=1}^K  (-1)^k Tr ( H^{(1)} f H^{(3)}  ( H^{(1)} H^{(3)})^{k-1} ) \prod_{\ell=1}^{K-k} \Bigg( \sum_{n_{\ell}=0}^\infty {(-)^{\ell n_{\ell}}\over n_{\ell}}  {\cal C}_{\ell}^{n_{\ell}}\Bigg)   \delta_{\sum_{\ell=1}^K \ell\, n_\ell\,;\,K-k} 
 \label{O3P8}
 \end{align}
where we have introduced the quantity
 \begin{equation}
{\cal C}_\ell = \frac{1}{\ell} Tr ( (H^{(1)} H^{(3)})^{\ell}) - \frac{3}{2}2\alpha'k{\cdot}H^{(1)}{\cdot}H^{(3)}{\cdot}k\,.
\label{O3P9}
\end{equation}
This expression is in agreement with the amplitude \eqref{eq:amplitudeB} computed in the manifestly covariant formalism. 

If we require the following additional restriction on the polarisations $H^{(1)}$ and $H^{(3)}$ :
\be
H^{[\mu\nu]}\,\Rightarrow \, H_{f}^{[\mu\nu]}= {1\over 2}(\rho^{\mu}\tilde{\rho}^{\nu}-\rho^{\nu}\tilde{\rho}^{\mu})
\ee
with constraints 
\be
\rho^{2}=\tilde{\rho}^{2}=\rho{\cdot}\tilde{\rho}=0\,,
\ee
one can see that the general contraction 
\be
((H^{(1)}H^{(3)})^{n})_{\mu\nu}\equiv H^{(1)}_{\mu\nu_{1}}H^{(3)\nu_{1}}_{\nu_{2}}...H^{(1)\nu_{n-1}}_{\nu_{n}}H^{(3)\nu_{n}}_{\nu}
\ee
reduces to 
\be
((H^{(1)}H^{(3)})^{n})^{\mu\nu}=\frac{1}{2^{n-1}}[\mathrm{Tr}(H_{f}^{(1)}H_{f}^{(3)})]^{n-1}(H_{f}^{(1)}H_{f}^{(3)})^{\mu\nu},
\ee
and also the amplitudes turn out to be described by a very compact expression 
\be
A(\{H_{f}^{(1)}\}_{K},{\cal E},\{H_{f}^{(3)}\}_{K})=\sqrt{2\alpha'}{\cal E}{\cdot}p_{3}\left(-{1\over 2}Tr(H_{f}^{(1)}H_{f}^{(3)})\right)^{K}\,L_{K}^{(1)}\left(3{2\alpha' k{\cdot}H_{f}^{(1)}{\cdot}H_{f}^{(3)}{\cdot}k\over Tr(H_{f}^{(1)}H_{f}^{(3)})}\right)+
\ee 
$$
-2\sqrt{2\alpha'}Tr(H_{f}^{(1)}f H_{f}^{(3)})\left(-{1\over 2}Tr(H_{f}^{(1)}H_{f}^{(3)})\right)^{K-1}\,L_{K-1}^{(2)}\left(3{2\alpha' k{\cdot}H_{f}^{(1)}{\cdot}H_{f}^{(3)}{\cdot}k\over Tr(H_{f}^{(1)}H_{f}^{(3)})}\right)
$$
which reproduces \eqref{3pt2HaZetfac}, computed in the manifestly covariant formalism, with the identification of $K$ with $s$.
.

\end{appendix}

\bibliographystyle{JHEP}
\bibliography{references}

\end{document}